\documentclass[prd,superscriptaddress,nofootinbib]{revtex4}
\usepackage{amsmath,amssymb,epsfig}
\usepackage[hypertex]{hyperref}
\usepackage{natbib,ifthen}

\begin{document}

\newcommand{\fixme}[1]{{\textbf{Fixme: #1}}}
\newcommand{\detD}{{\det\!\cld}}
\newcommand{\clh}{\mathcal{H}}
\newcommand{\ud}{{\rm d}}
\renewcommand{\eprint}[1]{\href{http://arxiv.org/abs/#1}{#1}}
\newcommand{\adsurl}[1]{\href{#1}{ADS}}
\newcommand{\ISBN}[1]{\href{http://cosmologist.info/ISBN/#1}{ISBN: #1}}
\newcommand{\jcap}{J.\ Cosmol.\ Astropart.\ Phys.}
\newcommand{\mnras}{Mon.\ Not.\ R.\ Astron.\ Soc.}
\newcommand{\progress}{Rep.\ Prog.\ Phys.}
\newcommand{\prlett}{Phys.\ Rev.\ Lett.}
\newcommand{\vort}{\varpi}
\newcommand\ba{\begin{eqnarray}}
\newcommand\ea{\end{eqnarray}}
\newcommand\be{\begin{equation}}
\newcommand\ee{\end{equation}}
\newcommand\lagrange{{\cal L}}
\newcommand\cll{{\cal L}}
\newcommand\cln{{\cal N}}
\newcommand\clx{{\cal X}}
\newcommand\clz{{\cal Z}}
\newcommand\clv{{\cal V}}
\newcommand\cld{{\cal D}}
\newcommand\clt{{\cal T}}

\newcommand\clo{{\cal O}}
\newcommand{\cla}{{\cal A}}
\newcommand{\clp}{{\cal P}}
\newcommand{\clr}{{\cal R}}
\newcommand{\uD}{{\mathrm{D}}}
\newcommand{\calE}{{\cal E}}
\newcommand{\calB}{{\cal B}}
\newcommand{\curl}{\,\mbox{curl}\,}
\newcommand\del{\nabla}
\newcommand\Tr{{\rm Tr}}
\newcommand\half{{\frac{1}{2}}}
\newcommand\fourth{{1\over 8}}
\newcommand\bibi{\bibitem}
\newcommand{\kf}{\beta}
\newcommand{\kfprod}{\alpha}
\newcommand\calS{{\cal S}}
\renewcommand\H{{\cal H}}
\newcommand\K{{\rm K}}
\newcommand\mK{{\rm mK}}
\newcommand\synch{\text{syn}}
\newcommand\opacity{\tau_c^{-1}}

\newcommand{\Psil}{\Psi_l}
\newcommand{\bsigma}{{\bar{\sigma}}}
\newcommand{\bI}{\bar{I}}
\newcommand{\bq}{\bar{q}}
\newcommand{\bv}{\bar{v}}
\renewcommand\P{{\cal P}}
\newcommand{\numfrac}[2]{{\textstyle \frac{#1}{#2}}}

\newcommand{\la}{\langle}
\newcommand{\ra}{\rangle}

\newcommand{\Omtot}{\Omega_{\mathrm{tot}}}
\newcommand\xx{\mbox{\boldmath $x$}}
\newcommand{\phpr} {\phi'}
\newcommand{\gam}{\gamma_{ij}}
\newcommand{\sqgam}{\sqrt{\gamma}}
\newcommand{\delk}{\Delta+3{\K}}
\newcommand{\dph}{\delta\phi}
\newcommand{\om} {\Omega}
\newcommand{\dom}{\delta^{(3)}\left(\Omega\right)}
\newcommand{\rar}{\rightarrow}
\newcommand{\Rar}{\Rightarrow}
\newcommand\gsim{ \lower .75ex \hbox{$\sim$} \llap{\raise .27ex \hbox{$>$}} }
\newcommand\lsim{ \lower .75ex \hbox{$\sim$} \llap{\raise .27ex \hbox{$<$}} }
\newcommand\bigdot[1] {\stackrel{\mbox{{\huge .}}}{#1}}
\newcommand\bigddot[1] {\stackrel{\mbox{{\huge ..}}}{#1}}
\newcommand{\Mpc}{\text{Mpc}}
\newcommand{\Al}{{A_l}}
\newcommand{\Bl}{{B_l}}
\newcommand{\eAl}{e^\Al}
\newcommand{\ix}{{(i)}}
\newcommand{\ixp}{{(i+1)}}
\renewcommand{\k}{\beta}
\newcommand{\HD}{\mathrm{D}}

\newcommand{\nonflat}[1]{#1}
\newcommand{\Cgl}{C_{\text{gl}}}
\newcommand{\Cgltwo}{C_{\text{gl},2}}
\newcommand{\He}{{\rm{He}}}
\newcommand{\Mhz}{{\rm MHz}}
\newcommand{\vx}{{\mathbf{x}}}
\newcommand{\ve}{{\mathbf{e}}}
\newcommand{\vv}{{\mathbf{v}}}
\newcommand{\vk}{{\mathbf{k}}}
\newcommand{\vn}{{\mathbf{n}}}

\newcommand{\vnhat}{{\hat{\mathbf{n}}}}
\newcommand{\vkhat}{{\hat{\mathbf{k}}}}
\newcommand{\taueps}{{\tau_\epsilon}}

\newcommand{\vgrad}{{\mathbf{\nabla}}}
\newcommand{\fbarln}{\bar{f}_{,\ln\epsilon}(\epsilon)}


\title{Testing General Relativity with 21\,cm intensity mapping}

\author{Alex Hall}
\email{ach74@ast.cam.ac.uk}
\affiliation{Institute of Astronomy and Kavli Institute for Cosmology, Madingley Road, Cambridge, CB3 0HA, U.K.}

\author{Camille Bonvin}
\affiliation{Institute of Astronomy and Kavli Institute for Cosmology, Madingley Road, Cambridge, CB3 0HA, U.K.}
 \affiliation{DAMTP, Centre
for Mathematical Sciences, Wilberforce Road, Cambridge CB3 0WA, U.K.}

\author{Anthony Challinor}
\affiliation{Institute of Astronomy and Kavli Institute for Cosmology, Madingley Road, Cambridge, CB3 0HA, U.K.}
 \affiliation{DAMTP, Centre
for Mathematical Sciences, Wilberforce Road, Cambridge CB3 0WA, U.K.}



\begin{abstract}
We investigate the prospects for constraining alternative theories of gravity with a typical near-term low-budget 21\,cm intensity mapping experiment. We derive the 21\,cm brightness temperature perturbation consistently in linear theory including all line-of-sight and relativistic effects. We uncover new terms that are a small correction on large scales, analogous to those recently found in the context of galaxy surveys. We then perform a Fisher matrix analysis of the $B_0$ parametrization of $f(R)$ gravity, where $B_0$ is proportional to the square of Compton wavelength of the scalaron. We find that our 21\,cm survey, in combination with CMB information from Planck, will be able to place a 95\% upper limit of $7 \times 10^{-5}$ on $B_0$ in flat models with
a $\Lambda$CDM expansion history, improving on current cosmological constraints by several orders of magnitude.
We argue that this constraint is limited by our ability to model the mildly non-linear regime of structure formation in modified gravity. We also perform a model-independent principal component analysis on the free functions introduced into the field equations by modified gravity, $\mu$ and $\Sigma$. We find that 20--30 modes of the free functions will be `well-constrained' by our combination of observables, the lower and upper limits dependent on the criteria used to define the `goodness' of the constraint. These constraints are found to be robust to uncertainties in the time-dependence of the bias. Our analysis reveals that our observables are sensitive primarily to temporal variations in $\Sigma$ and scale variations in $\mu$. We argue that the inclusion of 21\,cm intensity maps will significantly improve constraints on any cosmological deviations from General Relativity in large-scale structure in a very cost-effective manner.
\end{abstract}

\maketitle

\section{Introduction}
\label{sec:intro}

A major challenge in cosmology is understanding the late-time accelerated expansion of the Universe~\citep{riess_98,perlmutter_99}.
If we assume that gravity is described by General Relativity (GR), such an expansion requires the existence of a new exotic form of energy
with negative pressure, called dark energy. 
An alternative solution is to give up GR at large (cosmological)
scales and introduce modifications to the laws of gravity (for a review, see~\citep{2012PhR...513....1C}).

Whilst observations that measure the expansion history can rule out
specific dark energy and modified-gravity models, there is usually sufficient freedom
in the form of the modification to reproduce any desired expansion
history, thus rendering these measurements unable to distinguish
between a (possibly time-dependent) dark energy component and modified
gravity \citep{2007PhRvD..75d4004S}. To constrain these models, we must therefore go beyond the level of the background, and study the perturbed
Universe. It is here that an unclustered
dark energy component may be distinguished from modified
gravity\footnote{Note however that some modified-gravity models may still remain indistinguishable from 
a clustered dark energy model at the perturbed level ~\citep{2007PhRvL..98l1301K,2009PhRvD..80l3001K}.}.

Current and future optical surveys focus mainly on two promising observables: galaxy number counts and cosmic shear. 
These observables are both sensitive to the distribution of matter in the Universe and therefore provide powerful probes of the laws of gravity. 
In this paper we concentrate on a complementary observable measured in radio surveys:  the redshifted 21\,cm emission line of neutral atomic hydrogen (HI). We are interested in the post-reionization emission that traces the distribution of hydrogen at low redshifts. The
advantage is that after reionization, the physics of the 21\,cm power spectrum is simple and can be accurately
modelled in linear theory \citep{MNR:MNR12568,MNR:MNR15019}. This is
in contrast to observations of the epoch of reionization, where the 21\,cm power spectrum is very sensitive to the details
of the reionization process~\citep{0004-637X-653-2-815}.

The post-reionization signal is thought to be dominated by
emission from dense self-shielded damped Ly$\alpha$ systems~(DLAs)
which form in high-density regions, with some of the signal
originating from optically-thin Ly$\alpha$ absorbers in low-density
regions (for reviews, see
\citep{0034-4885-75-8-086901,2010ARA&A..48..127M,2006PhR...433..181F}).
On large (linear) scales, it is expected that these sources trace out
the underlying matter density field, up to a scale-independent bias factor if
the underlying density field is Gaussian (for corrections to this argument from primordial non-Gaussianity, see for example \citep{2008PhRvD..77l3514D}). 
The 21\,cm post-reionization signal therefore provides a new window to observe the distribution of matter.
Recently, measurements of the 21\,cm cosmological signal at $z=0.8$ have been
made by
cross-correlating with optical galaxies~\citep{2009MNRAS.394L...6P, 2012arXiv1208.0331M}.

The main advantage of the 21\,cm method over standard optical surveys is that it does not require the resolution of individual galaxies:
one detects the diffuse line radiation from a large number of sources.
This technique, called \emph{21\,cm intensity mapping}~\citep{2009astro2010S.234P} thus allows one
to observe large volumes of the sky efficiently
with low-resolution interferometers or single-dish experiments which can be constructed relatively cheaply
\citep{2012arXiv1209.1041B, CHIME, 0004-637X-721-1-164}. 
It also permits observation at higher redshifts than optical surveys, which are limited by the necessity to identify galaxies individually, a task
that becomes increasingly difficult as galaxies appear fainter.
 A further advantage of observing the 21\,cm line is that source redshifts are readily obtainable from the
observation frequency. This allows one to perform a tomographic analysis with very thin redshift bins, and consequently to
study the evolution and scale dependence of the underlying matter density field. 
Moreover, precise redshift information enhances the significance of the redshift-space distortion term, which is 
suppressed in optical surveys with only photometric redshifts by
uncertainties in the redshifts.
It follows that 21\,cm surveys are very well suited to
cosmological parameter inference \citep{2009JCAP...10..030V},
measurement of baryon acoustic oscillations
(BAO)~\citep{2010PhRvD..81j3527M,2012ApJ...752...80M}, constraining the
properties of dark energy \citep{2008PhRvL.100i1303C}, and
constraining models of modified gravity~\citep{2010PhRvD..81f2001M,2012arXiv1207.1273B}.

Experimentally, observing at low redshifts (i.e. after reionization) has the advantage that the
foreground contamination from Galactic synchrotron emission is
significantly reduced compared to epoch-of-reionization experiments,
since its amplitude scales roughly as $\nu^{-2.7}$ (for a thorough
investigation of foregrounds in the context of a single dish
experiment, see \citep{2012arXiv1209.1041B}). Although foreground
removal will still present a significant challenge in recovering the
cosmological signal, this should be possible since the foreground
components are smooth in frequency space, as opposed to the signal
which has significant structure in the frequency (i.e. radial)
direction due to the discrete nature of the underlying sources and their
small-scale clustering.

However, there is a price to pay for observing the Universe at high
frequencies. Since most of the HI is destroyed during the epoch of
reionization, the signal is significantly smaller at post-reionization
redshifts. Although the foregrounds are also smaller as mentioned
above, the system temperature of the detector is still significant due
to the receiver noise floor. A typical antenna temperature for a
near-term experiment is 50\,K , while the sky temperature at low
redshifts is typically 5--10\,K~\citep{2010ARA&A..48..127M}. The system
temperature at the receiver is thus dominated by the design of the
antenna.
A further disadvantage of working with the low redshift signal is that
the typical co-moving length scale below which linear perturbation
theory is expected to break down is larger than at high redshift, due
to hierarchical structure formation. Since we do not attempt to model
the non-linear 21\,cm signal in this work, the number of independent
modes of the density field accessible to us is much smaller than at
high redshift.
However, even with these considerations, the 21cm intensity mapping method provides a potentially powerful tool for learning about late-time structure formation, at a fraction of the cost of spectroscopic galaxy surveys or high-redshift epoch-of-reionization 21\,cm surveys such as those planned with the Square Kilometre Array (SKA).

In this paper, we provide a consistent derivation of
the perturbation to the 21\,cm brightness temperature in linear theory,
including all relativistic effects which could be relevant on large
angular scales, analogous to the new terms found in the context of
galaxy surveys in \citep{2011PhRvD..84f3505B,2011PhRvD..84d3516C,2009PhRvD..80h3514Y}. 
Our forecasts of the potential of 21\,cm mapping for constraining
modified-gravity models complements the work of~\citet{2010PhRvD..81f2001M},
which considered only information from the BAO signal and the
reconstructed weak lensing signal in two specific models. In contrast,
we include the full
three-dimensional clustering information in angular and redshift space
provided by the intensity maps and consider more general modifications
to gravity. We
try to keep our experimental setup as general as possible, but choose
a redshift range of $0.7 \leq z \leq 2.5$, appropriate to that of the
Canadian Hydrogen Intensity Mapping Experiment (CHIME)~\citep{CHIME}.

The plan of this paper is as follows. In Sec.~\ref{sec:21cm} we derive the mean 21\,cm brightness temperature and its perturbation consistently in linear theory, including all line-of-sight and relativistic effects. In Sec.~\ref{sec:MG} we discuss models of modified gravity, and the response of the 21\,cm signal to deviations from GR. We also set up the parametrizations of modifications to gravity there which are used for our forecasts. In Sec.~\ref{sec:IM} we discuss 21\,cm intensity mapping and the experimental setup assumed, before describing our statistical methods and presenting our results in Sec.~\ref{sec:methods}. We conclude in Sec.~\ref{sec:conc}.

The cosmological parameters varied in our statistical analyses are $(\Omega_bh^2,\Omega_ch^2,\tau,h,A_s,n_s,\Omega_{\nu}h^2)$, with fiducial values of $(0.0223,0.112,0.085,0.73,2.04\times10^{-9},0.966,0.0006)$ and pivot scale $0.05\,
\text{Mpc}^{-1}$ for the primordial power spectrum.
The background expansion history is always that of $\Lambda$CDM models, and we assume three neutrino species with degenerate masses.
Our metric convention is $+,-,-,-$ and,
unless stated otherwise, we take $c=1$.

\section{The 21\,cm brightness temperature}
\label{sec:21cm}

In this section we derive the perturbation to the 21\,cm brightness
temperature in linear theory, including all relativistic effects. Much
of the material in this section is drawn from
\citep{2007PhRvD..76h3005L} and \citep{2011PhRvD..84d3516C}. 
The main difference with respect to~\citep{2011PhRvD..84d3516C} is that here we are interested in
the relativistic contributions to the temperature rather than to the galaxy number counts.
Except where stated explicitly, the results in this section are valid for
all metric theories of gravity.


Let the rest-frame (proper) number density of neutral hydrogen atoms
at redshift $z$ along some line of sight be $n_{\text{HI}}$, with a fraction
$n_1/n_{\text{HI}}$ being in the excited triplet states and $n_0/n_{\text{HI}}$ in the
singlet state of the 21\,cm hyperfine transition. 
Neglecting the finite line width of the emission, which should be fine on the large scales we consider, in the rest-frame of the gas the net change in the
number of photons per volume
with energy between $E$ and $E + \ud E$ propagating within a solid angle $\ud \Omega$
in proper time $\ud t$ due to 21\,cm interactions is
\begin{equation}
\ud n_{\text{emit}} = \frac{1}{4\pi}\left[(n_1 - 3n_0)\mathcal{N}_{\gamma} + n_1\right]A_{10}\delta(E-E_{21})\ud E\ud t \ud \Omega .
\label{eq:nphotemit}
\end{equation}
Here, $\mathcal{N}_{\gamma}$ is the photon occupation number, 
$A_{10} \approx 2.869 \times 10^{-15}\; \mathrm{ s^{-1}}$
is the spontaneous emission coefficient~\citep{1952ApJ...115..206W},
and $E_{21} = 5.88 \, \mu\mathrm{eV}$
is the rest-frame energy of a 21\,cm photon.
In Eq.~\eqref{eq:nphotemit}, we have assumed that
the atomic triplet states are populated isotropically (see~\cite{2007PhRvD..76h3005L} for further discussion of this point).

The level populations define the spin temperature $T_s$ by $n_1/n_0 = 3e^{-T_{21}/T_s}$, where $T_{21} \equiv E_{21}/k_{B} = 0.068 \,\mathrm{K}$ and $k_{B}$ is Boltzmann's
constant. We assume that the radiation field at the relevant frequencies
consists of the CMB
blackbody with temperature $T_{\rm CMB} \gg T_{21}$ and the additional photons
from the 21\,cm interaction. At low redshift, $T_s \gg T_{\rm CMB}$~\citep{0034-4885-75-8-086901} since the spin temperature is coupled to the gas temperature by Ly$\alpha$ photons, and
stimulated emission and absorption can be neglected\footnote{This follows
for the CMB photons since $T_s \gg T_{\rm CMB}$. For the additional 21\,cm
photons, the optical depth for re-absorption can be shown to
be $O(10^{-2})T_{21}/T_s \ll 1$ around $z=1$, and is therefore negligible.}.
In this limit, Eq.~\eqref{eq:nphotemit} becomes
\begin{equation}
\ud n_{\text{emit}} \approx \frac{3}{16\pi} n_{\text{HI}} A_{10}
\delta(E-E_{21}) \ud E\ud t \ud \Omega \, .
\label{eq:nphotemit2}
\end{equation}
%
Note that this result is independent of the spin temperature.

Since re-absorption is negligible, and neglecting Thomson scattering of
the anisotropies in the line radiation (the Thomson optical depth to
$z=2$ is $0.008$), we can calculate the brightness temperature
by simply summing up the emitted photons.
These are received by an observer with 4-velocity $u^a$
along a line-of-sight direction $\vnhat$. In terms of the photon
distribution function $f(E,\vnhat)$, the number of photons collected with
energies between $E$ and $E+\ud E$ in an area $\ud A$ subtending a
solid angle at the observer of $\ud \Omega$ in proper time $\ud t$ is
\begin{equation}
\ud n_{\text{rec}} = f(E,\vnhat)E^2\ud E\ud \Omega\ud A\ud t .
\label{eq:nphotrec}
\end{equation}
We relate $\ud n_{\text{rec}}$ and $\ud n_{\text{emit}}$ by considering the
propagation of the congruence of null geodesics that focus at the
observer. Let $\ud \tilde{A}$ be the invariant area of
the wavefront at affine parameter $\lambda$ corresponding to some
source position. In an interval of affine parameter $\ud \lambda$, the
wavefront sweeps out a volume $\ud \tilde{A}u_s^a k_a \ud \lambda$,
where $u_s^a$ is the source 4-velocity and $k^a$ is the wavevector,
$k^{a} = \ud x^a/\ud \lambda$. If
the collecting area of the detector, $\ud A$, subtends a solid angle $\ud
\tilde{\Omega}$ at the source in its rest-frame, then photons
in an (observer-frame) energy range $\ud E$ around energy $E$ will be
collected in time $\ud t$ in solid angle $\ud \Omega$ with number
\begin{eqnarray}
\ud n_{\text{rec}} &=& \frac{3}{16\pi} \int \, \ud \lambda \left[
n_{\text{HI}} A_{10} \delta [E(1+z)-E_{21}] (1+z)\ud E \frac{\ud t}{1+z}
\ud \tilde{\Omega} \ud \tilde{A} k_a u^a_s\right] \nonumber \\
&=&  \frac{3}{16\pi} n_{\text{HI}} A_{10} (1+z) 
\left | \frac{\ud \lambda}{\ud z} \right | \ud E \ud t \ud \tilde{\Omega}
\ud \tilde{A} ,
\label{eq:nphotrec2}
\end{eqnarray}
where the integral in the first equation is along the line of sight, and
in the second the quantities are evaluated at redshift $z$ along the 
line of sight, where $1+z = E_{21} / E$. Equation~(\ref{eq:nphotrec2}) follows from integrating the product of $\ud n_{\text{emit}}$ from Eq.~(\ref{eq:nphotemit2}) and the
volume element $\ud \tilde{A}u_s^a k_a \ud \lambda$ along the line of sight,
and, additionally, we have used
$k_a u^a_s = E_{21}$. Using
the reciprocity relation $\ud \tilde{A}\ud \tilde{\Omega} = \ud A\ud
\Omega/(1+z)^2$, comparison with Eq.~\eqref{eq:nphotrec} gives
\begin{equation}
f(E,\vnhat) = \frac{3}{16\pi} \frac{n_{\text{HI}} A_{10} (1+z)}{E_{21}^2}
\left | \frac{\ud \lambda}{\ud z} \right |.
\label{eq:photdist}
\end{equation}
Finally, the 21\,cm brightness temperature $T_b$ is related to the
photon distribution function by $k_B T_b = h_p^3Ef/2$, where $h_p$ is
Planck's constant. It follows that
\begin{equation}
T_b(z,\vnhat) = \frac{3}{32\pi} \frac{h_p^3 n_{\text{HI}} A_{10}}{k_B E_{21}}
\left | \frac{\ud \lambda}{\ud z} \right |.
\label{eq:Tb_exact}
\end{equation}

If we initially ignore perturbations, $| \ud z / \ud \lambda| = 
(1+z) H(z) E_{21}$ (where $H(z)$ is the Hubble parameter)
and the background brightness temperature is (reinstating
factors of $c$)
\begin{eqnarray}
\bar{T}_b(z) &=& \frac{3 (h_p c)^3 \bar{n}_{\text{HI}} A_{10}}{32\pi k_B E_{21}^2
(1+z)H(z)} \nonumber \\
&=& 0.188\, \text{K} \, h \Omega_{\text{HI}}(z) \frac{(1+z)^2}{E(z)}  ,
\label{eq:Tbackground}
\end{eqnarray}
where $E(z) = H(z)/H_0$, the Hubble constant $H_0 = 100 h\,\text{km}
\text{s}^{-1} \text{Mpc}^{-1}$, and $\Omega_{\text{HI}}(z)$ is the \emph{comoving} mass density in HI in units of the \emph{current} critical density. Where
required in this work, we take $\Omega_{\text{HI}}= 4\times10^{-4}$ to be constant,
consistent with the local value found in the HIPASS survey \citep{2005MNRAS.359L..30Z}. The value of $\Omega_{\text{HI}}$ is not expected to vary significantly over the redshift range we consider \citep{2009ApJ...696.1543P}.

In the presence of perturbations, in Eq.~\eqref{eq:Tb_exact} we must evaluate
the perturbed $n_{\text{HI}}$ at the perturbed position appropriate to
the redshift $z$ and line of sight $\vnhat$, and include the perturbation
in $| \ud z / \ud \lambda|$. We will work in the
conformal Newtonian gauge, where the metric takes the form
\begin{equation}
ds^2 = a^2(\eta)\left[(1+2\Psi)d\eta^2 - (1-2\Phi)\delta_{ij}dx^idx^j \right].
\label{eq:metric}
\end{equation}
Here, $\eta$ is conformal time, $a(\eta)$ is the scale factor and the
spacetime-dependent gravitational potentials are $\Psi$ and $\Phi$.

An observer at rest in this coordinate system has 4-velocity $u^{\mu}
= a^{-1}(1 - \Psi)\delta_0^{\mu}$, which is parallel to
$\partial_{\eta}$. We can equip this observer with an orthonormal
tetrad of 4-vectors, $(e_{\alpha})^{a}$ such that $(e_0)^{a} = u^{a}$
and $e_i = a^{-1}(1 + \Phi)\partial_i$.
Decomposing the wavevector onto the tetrad gives
\begin{equation}
\frac{\ud \vx}{\ud\eta} = (1+\Phi + \Psi) \ve , \qquad
\frac{\ud\eta}{d\lambda} = a^{-2} \epsilon (1-\Psi),
\label{eq:kdecomp}
\end{equation}
where the comoving energy $\epsilon$ is defined by $k_a u^a \equiv
\epsilon/a$, and the unit 3-vector $\ve$ consists of the spatial tetrad
components of the propagation direction.
In addition we have the geodesic equations for the photons
\begin{equation}
\frac{\ud \ve}{\ud\eta} = - \vgrad_{\perp} (\Phi + \Psi), \qquad
\frac{\ud \epsilon}{\ud \eta} = - \epsilon
\frac{\ud \Psi}{\ud \eta} + \epsilon (\dot{\Phi} + \dot{\Psi}),
\label{eq:5}
\end{equation}
where $\vgrad_\perp \equiv \vgrad - \ve \ve \cdot \vgrad$. The
derivatives here are along the line of sight and overdots
denote partial derivatives with respect to $\eta$.

The source velocity (i.e.\ the bulk velocity of the HI) may be written as
$u_s^a = u^a + v^a$ such that
$u^\mu_s = a^{-1} [1-\Psi, v^i]$ where $v^i$ are the spatial tetrad
components of $v^a$. Similarly, the 4-velocity of
an observer at position $A$ moving with 3-velocity $v^i_{oA}$ in the
conformal-Newtonian frame can be written $u^\mu_{oA} = a_A^{-1}
[1-\Psi_A, v^i_{oA}]$. The redshift of the source
measured by this observer at $A$ is then
\begin{equation}
1+z = \frac{a_A }{a }\frac{\epsilon}{\epsilon_A} \bigl(1+\vnhat \cdot [\vv-\vv_{o A}]\bigr),
\label{eq:7}
\end{equation}
where $\vnhat=-\ve$ is the line-of-sight direction of the photons as seen
by the observer. (Aberration and evolution of $\ve$ can be neglected
to the order we are working.)
Integrating Eq.~\eqref{eq:5} gives the ratio of Newtonian-gauge
energies
\begin{equation}
\frac{\epsilon}{\epsilon_A} = 1 + \Psi_A - \Psi + \int_{\eta_A}^\eta
(\dot{\Phi}+\dot{\Psi})\, \ud \eta' ,
\label{eq:8}
\end{equation}
and so the redshift is
\begin{equation}
1+z = \frac{a_A}{a(\eta)} \left(1+ \Psi_A - \Psi + \int_{\eta_A}^\eta
(\dot{\Phi} +\dot{\Psi})\, \ud \eta' + \vnhat \cdot [\vv-\vv_{oA}] \right) .
\label{eq:9}
\end{equation}
Writing the perturbed conformal time at redshift $z$ along the line of sight
$\vnhat$ as $\eta(\vnhat,z) = \bar{\eta}_z + \delta \eta$, where
$\bar{\eta}_z$ is the unperturbed value, it follows from Eq.~\eqref{eq:9}
that
\begin{equation}
\clh(\bar{\eta}_z) \delta \eta = \Psi_A - \Psi + \int_{\eta_A}^{\bar{\eta}_z}
(\dot{\Phi} +\dot{\Psi})\, \ud \eta' + \vnhat \cdot [\vv-\vv_{oA}],
\label{eq:10}
\end{equation}
where all quantities on the right-hand-side are evaluated on the
zero-order lightcone. Here, $\mathcal{H}\equiv \dot{a}/a$ is the conformal
Hubble parameter.

Writing the perturbed $n_{\text{HI}}$ as $\bar{n}_{\text{HI}}(1+\delta_n)$
and evaluating $\bar{n}_{\text{HI}}$ at $\eta(\vnhat,z)$, the brightness temperature in Eq.~\eqref{eq:Tb_exact} becomes
\be
T_b(z,\vnhat) = \frac{3 h_p^3 A_{10}}{32\pi k_B E_{21}}
\bar{n}_{\text{HI}}(\bar{\eta}_z)\left(1+
\delta_n + \frac{\dot{\bar{n}}_{\text{HI}}}{\bar{n}_{\text{HI}}}\delta \eta\right)\left | \frac{\ud \lambda}{\ud z} \right |.
\label{eq:Tb_expand}
\ee
For $| \ud \lambda / \ud z|$, we differentiate Eq.~\eqref{eq:9} and use
Eq.~\eqref{eq:kdecomp}; evaluating at the perturbed conformal time gives
\begin{eqnarray}
\left| \frac{\ud \lambda}{\ud z} \right|(z,\vnhat) &=& \frac{a(\bar{\eta}_z)}
{\clh(\bar{\eta}_z)E_{21}(1+z)} \left[1-\left(\frac{\dot{\clh}}{\clh}-\clh
\right)\delta \eta - \frac{1}{\clh}\frac{\ud\Psi}{\ud\eta} + \frac{1}{\clh}
(\dot{\Phi}+\dot{\Psi})
+ \frac{1}{\clh}\vnhat \cdot \frac{\ud \vv}{\ud \eta}
+ \Psi + \vnhat\cdot \vv \right] .
\label{eq:19}
\end{eqnarray}
%
%
Combining these results, and using $\mathcal{H} = a H$,
gives the background brightness temperature in Eq.~\eqref{eq:Tbackground}. 
%
Inserting Eqs.~\eqref{eq:10} and \eqref{eq:19} into Eq.~\eqref{eq:Tb_expand} we find that the fractional perturbation to the brightness
temperature is\footnote{We retain the local dipole term at the observer
for completeness, but, since this is observer-dependent at linear order,
we only consider multipoles $l \geq 2$ in our later forecasts.}
\begin{equation}
 \Delta_{T_b}(z,\vnhat) = \delta_n + \frac{\dot{\bar{n}}_{\text{HI}}}{n_{\text{HI}}}\delta\eta - \left(\frac{\dot{\clh}}{\clh} - \clh \right)\delta\eta - \frac{1}{\clh}\frac{\ud\Psi}{\ud\eta} + \frac{1}{\clh}(\dot{\Phi}+\dot{\Psi})+\frac{1}{\clh}\vnhat \cdot \frac{\ud \vv}{\ud \eta} +  \vnhat\cdot \vv + \Psi .
\label{eq:perturb}
\end{equation}
If the comoving number density of $\text{HI}$ is conserved at low redshift (i.e.\
the ionized fraction of hydrogen is constant),
$\dot{\bar{n}}_{\text{HI}}/\bar{n}_{\text{HI}} = -3\clh$.
Using the Euler equation
$\dot{\vv} + \clh \vv + \vgrad \Psi=0$,\footnote{Note that this equation is still valid in the modified-gravity models we are interested in.} some further
cancellations occur leaving
\begin{equation}
 \Delta_{T_b}(z,\vnhat) = \delta_n - \frac{1}{\clh}\vnhat \cdot (\vnhat \cdot
\vgrad \vv) + \left(\frac{\ud \ln (a^3 \bar{n}_{\text{HI}})}{\ud \eta}-
\frac{\dot{\clh}}{\clh} - 2\clh \right)\delta\eta + \frac{1}{\clh}\dot{\Phi} + \Psi .
\label{eq:perturb2}
\end{equation}
Equation~\eqref{eq:perturb2}, new to this work, is
the expression we have been seeking for the 21\,cm brightness
temperature perturbation in the limit of zero frequency bandwidth.
It is straightforward to show that it agrees with the more complete
expression [Eq.~(18)] in~\cite{2007PhRvD..76h3005L} in the limit
$T_s \gg T_{\text{CMB}}$ and for low optical depth for re-absorption
and Thomson scattering. This expression is the analogue of the expression in~\citep{2011PhRvD..84f3505B, 2011PhRvD..84d3516C}, but here applied to the 21\,cm brightness temperature rather than galaxy number counts.
As discussed in more detail below, the exact form of the relativistic terms differs from that for galaxy number counts
since the brightness temperature is also affected by perturbations in the luminosity distance.

Each of the terms on the right of Eq.~\eqref{eq:perturb2}
has a simple physical explanation. The first two
terms are the usual density and redshift-space distortion terms.
The third term comes from evaluating the
zero-order brightness temperature at the perturbed time corresponding
to the observed redshift. The time perturbation given in
Eq.~\eqref{eq:10} contains integrated Sachs-Wolfe (ISW), potential
and Doppler terms. The fourth term arises from the part of the ISW
term in Equation~\eqref{eq:9} that is not cancelled by the velocity
evolution introduced via the Euler equation. The final potential term
comes from the conversion between increments in redshift and radial
distance in the gas frame.

The perturbation to the brightness temperature depends only on the
observer velocity through $\delta \eta$. Isolating these terms gives
\begin{equation}
\Delta_{T_b}^{\vv_{oA}}(z,\vnhat) = \vnhat \cdot \vv_{oA} + 
\vnhat\cdot \vv_{oA} (1+z) \frac{\ud \ln \bar{T}_b}{\ud z} .
\label{eq:vterms}
\end{equation}
We compare this dependence with the exact result for the variation in
the brightness temperature under a change of frame:
$(1+z')T_b'(z',\vnhat') = (1+z)T_b(z,\vnhat)$ which follows from the
invariance of the distribution function. Since $1+z = (1+z')(1+\vnhat
\cdot \vv_{\text{rel}})$ to linear order, where $\vv_{\text{rel}}$ is the
relative velocity of the two observers, Taylor expanding the redshift
dependence of $T_b(z,\vnhat)$ about $z'$ we have
\begin{equation}
\Delta_{T_b}'(z',\vnhat') \approx \Delta_{T_b}(z',\vnhat) +
\vnhat \cdot \vv_{\text{rel}} + \vnhat \cdot \vv_{\text{rel}} (1+z')
\frac{\ud \ln \bar{T}_b}{\ud z'} ,
\end{equation}
which is clearly consistent with the velocity dependence in
Eq.~\eqref{eq:vterms}.

The large-scale clustering of the HI gas follows that
of the discrete sources (e.g.\ DLAs) housing the neutral gas. We assume
linear scale-independent bias of these sources in the comoving gauge,
following the arguments in
\citep{2011PhRvD..84d3516C}. During matter domination, the comoving
gauge coincides with the synchronous gauge, and we have
\begin{equation}
\delta_n = b\delta_m^{\text{syn}} +  \left( \frac{\ud \ln (a^3
\bar{n}_{\text{HI}})}{\ud \eta} - 3 \mathcal{H}\right)\frac{v}{k} ,
\label{eq:bias}
\end{equation}
where $v$ is the Newtonian-gauge matter/gas velocity with
$\vv = - k^{-1} \vgrad v$, and $\delta_m^{\text{syn}}$ is the synchronous-gauge
matter overdensity.

Equation~\eqref{eq:perturb2} can be compared with the related expression for the
differential number counts of discrete objects detailed in~\citep{2011PhRvD..84f3505B,2011PhRvD..84d3516C,2009PhRvD..80h3514Y,2012PhRvD..85b3504J}.
For simplicity, consider counting the HI atoms per solid angle and per
redshift $n^{\text{obs}}_{\text{HI}}(z,\vnhat)$.
The brightness temperature, Eq.~\eqref{eq:Tb_exact}, may be rewritten as 
\begin{equation}
T_b(z,\vnhat) = \frac{3 h_p^3 A_{10}}{32\pi k_B E_{21}^2} \frac{n^{\text{obs}}_{\text{HI}}(z,\vnhat)}{\text{det} \mathcal{D}} ,
\label{eq:altform}
\end{equation}
where $\text{det}\mathcal{D}$ is the determinant of the
Jacobi map in the observer's frame~\citep{1992grle.book.....S} which relates
the (Lorentz-invariant) area of a bundle of light rays to the solid angle at
the observer. The determinant of the Jacobi map is related to the square of the
luminosity distance $d_L$ through reciprocity: $d_L^2 = (1+z)^4 \text{det}
\mathcal{D}$. The presence of $d_L^2$ in Eq.~\eqref{eq:altform}
makes intuitive sense since we measure
surface brightness (i.e.\ angular density of sources multiplied by flux)
in 21\,cm intensity mapping experiments rather than counting
objects.
At linear order the fractional perturbation to the brightness temperature $\Delta_{T_b}(z,\vnhat)$ is then
related to the HI angular number density fluctuation $\Delta(z,\vnhat)$ (calculated following the galaxy counts calculation in~\citep{2011PhRvD..84f3505B,2011PhRvD..84d3516C,2009PhRvD..80h3514Y,2012PhRvD..85b3504J})  by
\be
\label{eq:linkgal}
\Delta_{T_b}(z,\vnhat)=\Delta(z,\vnhat)-2\frac{\delta d_L(z,\vnhat)}{\bar{d}_L(z)},
\ee
where $\bar{d}_L$ is the background luminosity distance and $\delta d_L$ denotes its first-order perturbation. 
The perturbations in the brightness temperature are therefore not only affected by the perturbations in the galaxy number count,
but also by the perturbations in the luminosity distance. The latter contains a lensing convergence term, a Doppler term and gravitational potential terms~\citep{2006PhRvD..73b3523B}. The precise form of the relativistic terms is therefore different for number counts and the brightness temperature.
In particular, Eq.~\eqref{eq:linkgal} tells us why the brightness temperature has no magnification term at
linear order, whereas the number count has one: the lensing convergence term in the luminosity distance exactly
cancels the one in the galaxy number count.
In other words, magnified sources give a greater observed flux but have reduced angular density on the sky and
lensing therefore conserves surface brightness.
Hence, as for the CMB, gravitational lensing has no effect at first order
to the 21\,cm brightness temperature: the lensed image of a smooth sky is a
smooth sky\footnote{Note, however, that at second order 21\,cm intensity mapping can be used to reconstruct the lensing power spectrum~\citep{2010PhRvD..81f2001M}.}.

For fixed redshift, the brightness temperature is a function on the sphere, that we can expand in spherical harmonics:
\be
\Delta_{T_b}(z,\vnhat)=\sum_{lm}\Delta_{T_b,lm}(z)Y_{lm}(\vnhat).
\ee
Expressing the perturbations in terms of their Fourier transforms, we can rewrite the 
$\Delta_{T_b,lm}(z)$ coefficients as
\begin{equation}
\Delta_{T_b,lm}(z) = 4\pi i^l \int \frac{\ud^3 \vk}{(2\pi)^{3/2}}
\Delta_{T_b,l}(\vk,z) Y_{lm}^*(\hat{\vk}).
\end{equation}
The first term in Eq.~\eqref{eq:perturb2} gives 
$\Delta_{T_b,l}(\vk,z) = \delta_n\, j_l(k \chi)$, where $\chi$ is the conformal distance
back to redshift $z$ in the background.
For the velocity term $\vv\cdot \vnhat$ we use that $\vv=-i\hat{\vk}v$, so that
$\vv\cdot \vnhat \exp(i\mathbf{k}\cdot \vnhat \chi)=-v\, \partial_{k\chi}\exp(i\mathbf{k}\cdot \vnhat \chi)$. This gives a contribution of the form
$\Delta_{T_b,l}(\vk,z) = -v\, j_l{}'(k \chi)$ where a prime denotes a derivative with respect to the argument. 
Similarly for the redshift-space distortion term we use the above identity twice. Putting everything together we find
\ba
\Delta_{T_b,l}(\vk,z)&=&\delta_n\, j_l(k \chi)+ \frac{kv}{\clh}j_l{}''(k\chi)+\left(\frac{1}{\clh}\dot{\Phi} + \Psi\right)j_l(k \chi)\\
&&-\left(\frac{1}{\clh}\frac{\ud \ln (a^3 \bar{n}_{\text{HI}})}{\ud \eta}-
\frac{\dot{\clh}}{\clh^2} - 2 \right)\left[\Psi\, j_l(k \chi)+v\, j_l{}'(k \chi) +\int_0^\chi (\dot{\Psi}+\dot{\Phi})j_l(k \chi')d\chi' \right],\nonumber
\ea
where we have neglected the contributions at the observer since they only give rise t monopole or dipole terms.

We then integrate over a redshift (frequency)
normalized window function $W(z)$ to give
\begin{equation}
\Delta_{T_b,l}^W(\vk) = \int_0^\infty \ud z W(z)
\Delta_{T_b,l}(\vk,z).
\label{eq:window}
\end{equation}
The angular cross-spectra between redshift windows is then calculated
as
\begin{equation}
C_l^{WW'} = 4\pi\int \ud \ln k \, \clp_{\clr}(k) \Delta_{T_b,l}^W(k) \Delta_{T_b,l}^{W'}(k),
\end{equation}
where $\clp_{\clr}(k)$ is the dimensionless power spectrum of the
primordial curvature perturbation $\clr$, and the transfer functions
$\Delta_{T_b,l}^W(k) \equiv \Delta_{T_b,l}^W(\vk)/\clr(\vk)$.

We calculate the cross-spectra
using a modified version of the Boltzmann code
\textsc{CAMB sources}\footnote{http://camb.info/sources.}.
At the low redshifts we consider, some care
must be taken when integrating over the narrow window function, and we
found it necessary to run the code at $\texttt{accuracy\_boost}=2$ to
ensure the window was well sampled.

The new terms that have been uncovered by this self-consistent linear
analysis are expected to be negligible on all but the largest scales
and highest redshifts (see~\citep{2011PhRvD..84f3505B} for a thorough
investigation in the context of galaxy surveys). The only new aspect
of our application is the narrow window function in
redshift-space which enhances the relative importance of the redshift-space
distortion term. Considering fluctuations at a given small angular scale $1/l$,
if the redshift window function is broad compared to the typical
linear scale $\chi(z)/l$ of the contributing perturbations (where
$\chi(z)$ is comoving distance back to redshift $z$), we are in the `Limber'
regime and the redshift distortions $\propto - \vnhat \cdot (\vnhat \cdot \vgrad
\vv)$ will tend to cancel out on integrating through the window. Normalising
the window function to unity to keep the integrated background $\bar{T}_b$
almost constant, the power from redshift-space distortions falls as $(\Delta
\chi)^{-2}$, where $\Delta \chi$ is the width of the window function, on small
scales. 
In the Limber limit, the power from the integrated density term falls more
slowly -- as $(\Delta \chi)^{-1}$ -- since the fluctuation \emph{power}
accumulates as the number
of incoherent patches within the window.
However, since the redshift
window is very narrow for 21\,cm mapping, only very small scales are in the
Limber regime, thus significantly enhancing the relative power in redshift-space
distortions, as seen in Fig.~6 of~\citep{2011PhRvD..84f3505B}.



\begin{figure}
\centering
\includegraphics[width=0.7\textwidth]{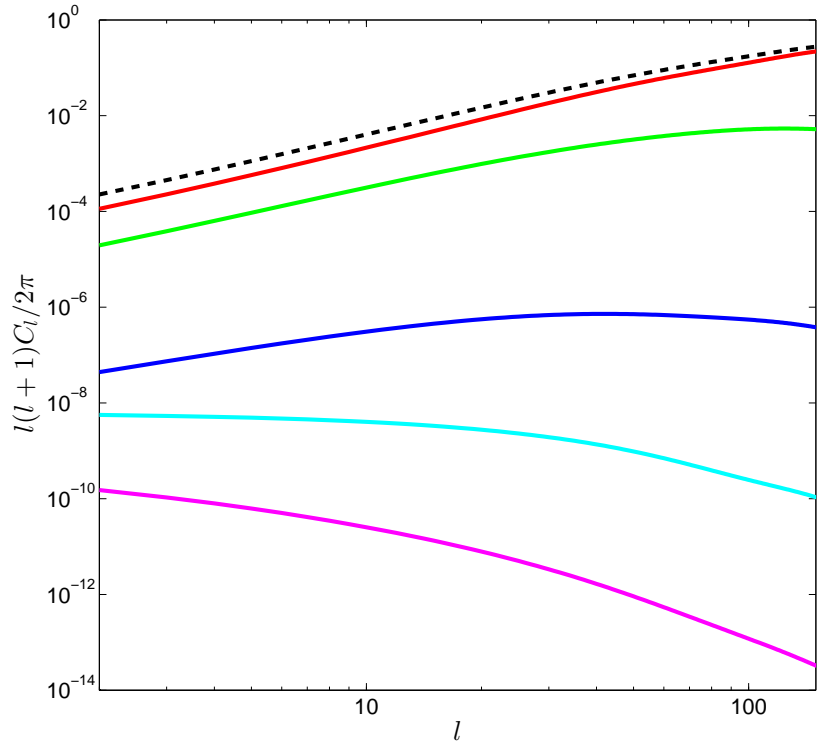}
\caption{(Color Online). Fractional brightness temperature perturbation power spectrum at $z=1$
  with a $2\,\mathrm{MHz}$ bandwidth. The auto-spectra of the full signal (black, dashed) and of each individual term in Eq.~\eqref{eq:perturb2} are shown, generically grouped (solid lines, top to bottom respectively) as Newtonian-gauge
  density (red), redshift-space distortions (green), velocity terms
  (blue), all potential terms evaluated at the source position (cyan)
  and the ISW term (magenta).}
\label{fig:newterms}
\end{figure}

\begin{figure}
\centering
\includegraphics[width=0.7\textwidth]{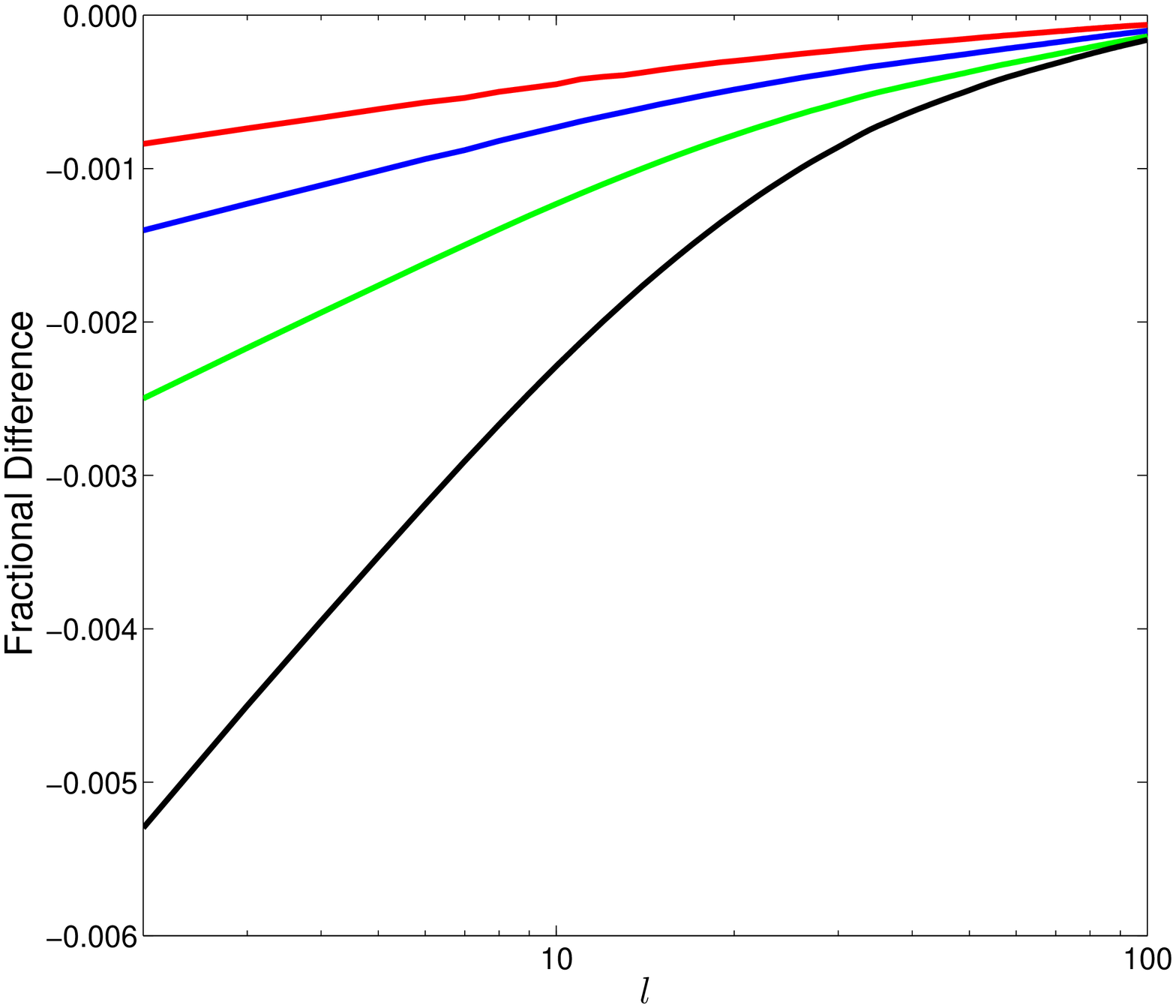}
\caption{(Color Online). Fractional error in the 21\,cm angular power spectrum
at $z=1$ if only density and redshift-space distortions terms are retained.
The errors are plotted for window functions of bandwidth (top to bottom respectively)
2\,MHz (red), 5\,MHz (blue), 10\,MHz (green), and 20\,MHz (black).}
\label{fig:newterms_binsize}
\end{figure}

In Fig.~\ref{fig:newterms} we plot the auto-spectra of each of the terms
in Eq.~\eqref{eq:perturb2} individually, including those contained in
$\delta \eta$, for a bandwidth of 2\,MHz at $z=1$. Clearly, the density and redshift-space distortion terms are dominant on all angular scales at this redshift.
Note that super-Hubble effects are generally suppressed in the 21\,cm power
spectrum since the signal on large angular scales is dominated not by modes
at the corresponding linear scale, but by smaller sub-Hubble scale modes~\cite{2007PhRvD..76h3005L}. This is because the dimensionless power spectrum of $\delta_n$ grows rapidly as $k^4$ on scales smaller than Hubble length (but larger than the horizon size at matter-radiation equality). The 21\,cm signal is therefore like white noise, $C_l = \mathrm{const.}$, on angular scales large compared to the angle subtended by the peak in the matter power spectrum.

The fractional error in the power spectrum if only the density\footnote{In the remainder of this work, `density' refers to the Newtonian-gauge density of Eq.~\eqref{eq:bias}.} and redshift-space
distortion terms are retained is shown in Fig.~\ref{fig:newterms_binsize}
at $z=1$ for various widths of redshift window.
We see that the relative importance of relativistic effects increases as the bin size increases, consistent with the results of \citep{2011PhRvD..84f3505B}.
This arises from the dominant white-noise contribution of small scale modes at a given $l$ being suppressed by cancellations through the width of the window, but the contribution of large-scale modes, where the relativistic terms are relevant, not being suppressed.
At the low redshifts considered in this work, the relativistic terms
represent only sub-percent corrections to the large-scale angular power, where
cosmic variance is large. Moreover, their effect is small compared to
astrophysical modelling uncertainties, for example in the bias.
However, we include the relativistic terms in our forecasts for consistency.


%
%
%
%
%
%
%
%


\section{Modified gravity}
\label{sec:MG}
Several attempts have been made to explain the phenomenon of
accelerating expansion through a modification to standard GR
\citep{2012PhR...513....1C}. These modifications must predict an
expansion history close to that of $\Lambda$CDM, but generically
predict different linear perturbation dynamics
\citep{2006JCAP...01..016K, 2007PhRvD..75d4004S,
  2009RPPh...72i6901S}. It is therefore necessary to study the
clustering of matter in order to distinguish between competing
theories\footnote{It should be borne in mind that none of these
  theories explains why the vacuum energy from particle physics is
  cancelled to such high precision on cosmological scales.}.
Note however that if dark energy is allowed to cluster, sufficiently complex models may 
be able to reproduce the perturbed dynamics of some modified-gravity models,
making them effectively indistinguishable~\citep{2007PhRvL..98l1301K,2009PhRvD..80l3001K}.

Amongst the most studied examples of modified gravity are
scalar-tensor theories and the higher-derivative theory $f(R)$, which can be mapped on to a scalar-tensor theory via a conformal transformation of the metric and a field redefinition \citep{2010RvMP...82..451S}. The
action in a scalar-tensor theory takes the form
\begin{equation}
S = \int \ud^4 x \sqrt{-g} \left[ -\frac{M_{\mathrm{Pl}}^2}{2}R +
  \frac{1}{2} ( \vgrad \phi )^2 - V(\phi) \right ] + S_m(\tilde{g}_{\mu
  \nu},\psi_m^{(i)}),
\label{eq:STaction}
\end{equation}
where $M_{\mathrm{Pl}} = 1/\sqrt{8\pi G}$ is the reduced Planck mass, $V$ is the potential for
 the scalar field $\phi$, and $\psi_m^{(i)}$ are the matter fields which
couple to the conformally rescaled metric $\tilde{g}_{\mu\nu}$ where
\begin{equation}
\tilde{g}_{\mu \nu} = e^{-\alpha(\phi)/M_{\text{Pl}}}g_{\mu \nu},
\label{eq:conf}
\end{equation}
where $\alpha(\phi)$ is an arbitrary coupling function.
The metric $g_{\mu \nu}$ is the Einstein-frame metric, where the
action in Eq.~\eqref{eq:STaction} looks like the standard
Einstein-Hilbert action but with matter non-minimally coupled to the
metric. This frame has the advantage of being mathematically `close'
to GR, but has the disadvantage that matter does not follow the
geodesics of $g_{\mu \nu}$ and the energy-momentum tensor is
not covariantly conserved with respect to this metric.
The frame picked out by $\tilde{g}_{\mu
  \nu}$ is the Jordan frame, where matter moves along geodesics but
the gravitational action is not of Einstein-Hilbert form.
Both frames are equivalent in the
sense that observables calculated in either will be the same. From now
on we will assume that all matter fields couple to the metric with a
universal coupling $\alpha(\phi)$.

We describe linear perturbations in the conformal Newtonian gauge. For
scalar modes, the field
equations reduce to four independent equations involving the gravitational potentials
$\Phi$ and $\Psi$, the fractional density perturbation $\delta$, the velocity potential $v$ (or equivalently the momentum density $q$), and the anisotropic stress
$\Pi$. The anisotropic stress is negligible in the late Universe, but
we include it in numerical work for consistency.
In the Jordan frame, energy-momentum is conserved and so the
continuity and Euler equations retain their forms in GR. In Fourier space, for
pressure-free matter, they become
\begin{equation}
\dot{\delta} +k v - 3\dot{\Phi} = 0, \qquad
\dot{v} + \clh v -k \Psi = 0,
\label{eq:bianchi}
\end{equation}
where the dots denote derivatives with respect to conformal time $\eta$. We use the Fourier convention
\be
\Psi(\vx, \eta)=\int \frac{d^3\vk}{(2\pi)^{3/2}} \Psi(\vk,\eta)e^{i \vk \cdot \vx},
\ee
and, recall, $v(\vk,\eta)$ is defined through $\mathbf{v} = -i \hat{\vk} v$. This leaves two independent equations that may receive
modifications. It is convenient to take these to be the 
Poisson equation, and the relationship
between the two potentials. In many models, including scalar-tensor theories,
the modifications on sub-Hubble scales (i.e.\ the quasi-static limit where
temporal derivatives are small compared to spatial derivatives) take the form
\begin{eqnarray}
-k^2 \Psi &=& 4 \pi G a^2 \mu(a,k) \bar{\rho} \Delta ,
\label{eq:MGPoisson} \\
\Phi &=& \gamma(a,k)\Psi,
\label{eq:MGslip}
\end{eqnarray}
where $\Delta$ is the comoving density perturbation $\Delta=\delta+3\mathcal{H}v/k$
and $\bar{\rho}$ is the background matter density (see, for example~\citep{2008JCAP...04..013A, 2010PhRvD..81j4023P}).
Equations~\eqref{eq:MGPoisson} and~\eqref{eq:MGslip} may be taken to define the functions $\mu$ and $\gamma$ which are generally functions of time and scale;
in GR, $\mu=\gamma=1$. Note that when the quasi-static limit is not obtained, we may still define $\mu$ and $\gamma$ through Eqs.~\eqref{eq:MGPoisson} and~\eqref{eq:MGslip} but these functions will now generally depend on the initial conditions.
Note that, for later convenience, we have written the modified Poisson equation
in terms of $\Psi$ whereas it involves $\Phi$ in GR.

It is straightforward to show that, despite the modifications in Eqs.~\eqref{eq:MGPoisson} and~\eqref{eq:MGslip}, the comoving curvature
perturbation $\mathcal{R} = - \Phi - \mathcal{H} v / k$ is conserved on
super-Hubble scales as it is in GR. Differentiating the expression for $\mathcal{R}$ and using Eqs.~\eqref{eq:bianchi}, \eqref{eq:MGPoisson} and \eqref{eq:MGslip} we find
\begin{equation}
\left[3 \frac{H'}{H} - \left(\frac{k}{aH}\right)^2\right] \mathcal{R}'
= \left(\frac{k}{aH}\right)^2 \left(\Phi'+\Psi\right) +
\frac{k^2 H'/H}{4\pi G a^2 \mu \bar{\rho}}
\left(\Psi' - [\ln (a^2 \mu \bar{\rho})]'\Psi\right) ,
\end{equation}
where $H = \mathcal{H}/a$ is the Hubble parameter and
primes denote derivatives with respect to $\ln a$. Since $\mathcal{R}
\sim \Phi$, the comoving curvature perturbation is conserved on super-Hubble scales
provided $\mu$ and $\gamma$ tend to scale-free functions there. Constancy
of the comoving curvature perturbation can be shown to hold very generally
for metric theories of gravity in which energy-momentum is conserved~\cite{2000PhRvD..62d3527W,2006ApJ...648..797B}. Differentiating $\mathcal{R}'=0$ gives a second-order
equation for the potentials which in our notation is
\begin{equation}
\Phi'' + \Psi' - \frac{H''}{H'} \Phi' + \left(\frac{H'}{H} - \frac{H''}{H'}\right) \Psi = 0
\qquad (k \ll a H) .
\end{equation}
Note that, once the relation between $\Phi$ and $\Psi$ is specified, the evolution of the
metric perturbations on super-Hubble scales is determined by the background
expansion history~\cite{2006ApJ...648..797B}. For matter-dominated expansion, $H' / H = H''/ H'$ and
$\Phi$ and $\Psi$ are constant on large scales if $\gamma$ is time-independent.
We are, of course, still free to modify the
relationship between $\Psi$ and $\Delta$; for example, $\mu>1$ on large scales with constant $\gamma$ would actually
\emph{suppress} large-scale clustering in $\Delta$ during matter-dominated expansion, despite increasing $\mu$ giving a larger effective Newton's constant.

On sub-horizon scales, we can rewrite the set of perturbation equations
as second-order evolution equations for the density and for the velocity.
Neglecting the $\dot{\Phi}$ term in the continuity equation and
replacing $\Delta$ with $\delta$ in the 
Poisson equation, we find
\ba
\ddot{\delta}+\H\dot{\delta}-4\pi G a^2 \mu \bar{\rho} \delta &=& 0
\label{eq:deltaevol},\\
\ddot{v} +\left(2\H-\frac{\dot{\mu}}{\mu}\right) \dot{v} +\left(
\dot{\H}+\H^2-\H \frac{\dot{\mu}}{\mu} - 4\pi G a^2 \mu \bar{\rho} \right)v&=&0.
\label{eq:vevol}
\ea
We see that the evolution of $\delta$ and $v$ on sub-horizon scales is completely determined by $\mu$. 
The density and redshift-space distortion terms in the 21\,cm brightness temperature are therefore insensitive on small scales to modifications in the $\gamma$ function. Note that this conclusion also applies to galaxy number counts. 
However $\delta$ and $v$ are expected to deliver stringent constraints on the time and scale dependence of $\mu$. 
In particular, since the evolution of the velocity is sensitive to both $\mu$ and its time derivative $\dot{\mu}$, it is strongly affected by sharp variation in $\mu$. The 21\,cm brightness temperature is therefore particularly well adapted to constrain $\mu$, thanks to its strong sensitivity to redshift-space distortions.

To understand qualitatively the effect of $\mu$ on the growth of perturbations on sub-Hubble scales, consider a toy-model with constant $\mu$ and matter-dominated
expansion. Equation~\eqref{eq:deltaevol} then becomes
\begin{equation}
\ddot{\delta} + \frac{2}{\eta} \dot{\delta} - \frac{6\mu}{\eta^2} \delta = 0 ,
\end{equation}
which has power-law solutions $\delta \propto a^{p_\pm}$ with
$p_\pm = (-1 \pm \sqrt{1+24\mu})/4$. As expected, $\mu > 1$ enhances
clustering and $\delta$ grows faster than in GR (where $\delta \propto a$)
and the gravitational potential wells of $\Psi$ deepen in time.

%

Although the above solutions were derived under the simplifying assumptions of a constant $\mu$ and sub-horizon scales during matter domination, the general effect of $\mu > 1$ on the clustering of matter is clear: a larger effective Newton's constant boosts the strength of gravity which enhances the degree of clustering on sub-Hubble
scales, thus allowing the gravitational potentials to grow with time. The time-dependence of $\Psi$ and $\Phi$ can source an ISW effect in the CMB temperature anisotropy, and alter the strength of gravitational lensing effects as a function of
redshift.

Since the models we consider attempt to explain the late-time
acceleration of the Universe, they only affect the linear dynamics at
late times when the acceleration sets in\footnote{There are exceptions, for example
$f(R)$ models with $\ud^2 f / \ud R^2 < 0$ have unstable perturbations at
high curvature for a standard expansion history~\cite{2007PhRvD..75d4004S}.}.
The CMB is thus only
sensitive to modified gravity at the perturbative level via
the large-scale (late-time) ISW effect, and the
CMB weak lensing signal. Both of these are sensitive to the Weyl
potential, which is equal to $(\Phi + \Psi)/2$ in the Newtonian gauge. The
Weyl potential is related to the matter density by
\begin{equation}
-k^2(\Phi + \Psi) = 4\pi G a^2 \mu (1+\gamma) \bar{\rho} \Delta = 8 \pi
 G a^2 \Sigma(a,k) \bar{\rho} \Delta,
\label{eq:Sigma}
\end{equation}%
where we have defined the new parameter
\begin{equation}
\Sigma \equiv \frac{\mu}{2}(1+\gamma).
\end{equation}
We will work in the ($\mu,\Sigma$) basis since the 21\,cm signal is
primarily sensitive to $\mu$ and the CMB is primarily sensitive to $\Sigma$,
although the CMB is also sensitive to $\mu$ through the
enhanced growth of $\Delta$ it induces.

Under fairly weak assumptions about the coupling function $\alpha(\phi)$ \citep{2009PhRvD..79h3513Z}, an action of the form of
Eq.~\eqref{eq:STaction} leads to the Bertschinger-Zukin~\cite{2008PhRvD..78b4015B} forms of $\mu$ and $\gamma$ in the Jordan frame given by
\begin{equation}
\mu(a,k) = \frac{1+\beta_1 \lambda_1^2 k^2 a^s}{1+\lambda_1^2 k^2 a^s},
\label{eq:BZmu}
\end{equation}
\begin{equation}
\gamma(a,k) = \frac{1+\beta_2 \lambda_2^2 k^2 a^s}{1+\lambda_2^2 k^2 a^s},
\label{eq:BZgamma}
\end{equation}
where $\beta_1$ and $\beta_2$ are coupling parameters, $\lambda_1$ and $\lambda_2$ are Compton wavelengths at the present epoch ($a=1$) with $s$ a constant describing their time evolution.
For scalar-tensor theories, only two of $\beta_i$ and $\lambda_i$ are
independent~\cite{2009PhRvD..79h3513Z}.
This form for $\mu$ and $\gamma$ has the advantage of being physically transparent; on scales much larger than the Compton wavelength of the scalar field $\mu$ and $\gamma$ obtain their GR forms, and on scales below the Compton wavelength the modifications are proportional to the coupling parameters $\beta_i$. In addition, for $s>0$, the comoving Compton wavelength grows with time, such that the influence of the scalar field perturbation can be confined to late times if desired.

The disadvantage of using the parametrizations in Eqs.~\eqref{eq:BZmu} and \eqref{eq:BZgamma} is that their domain of applicability is limited to scalar-tensor theories, or theories conformally equivalent. A more general, model-independent approach is desirable. The works of Baker et al. \citep{2011PhRvD..84l4018B, 2012arXiv1209.2117B} provide just such a model-independent formalism, but the large number of free functions present in their formalism is undesirable for our simple application here.

We argue that the most interesting theory that may be probed by upcoming large-scale structure surveys is the \emph{null} theory, i.e. we seek to test whether GR is correct on cosmological scales. To establish the veracity of this statement is a far more interesting challenge than choosing between the many alternative theories on offer, few of which have a compelling observational justification for pursuit.
To this end, we will perform a principal component analysis (PCA) of the functions $\mu$ and $\Sigma$. Since GR predicts that these functions are equal to unity at all times and (linear) scales, any evidence to the contrary would be particularly interesting. A study of the well-constrained functional forms provides a model-independent way of forecasting future constraints\footnote{However, since the number of PCA nodes is finite, we are still vulnerable to theories which may mimic GR at the nodes, but the use of a fine enough pixelization should mitigate this risk considerably.}. For applications
of PCA in testing modified gravity, see e.g.~\cite{2009PhRvL.103x1301Z,2010PhRvD..81j3510Z,2012PhRvD..85d3508H}. We outline the details of our PCA procedure in Sec.~\ref{sec:methods}.

We note in closing that the parametrization of Eqs.~\eqref{eq:BZmu} and \eqref{eq:BZgamma} is particularly ill-suited to inference when the maximum-likelihood point is close to GR. In this case, one of the parameters $(\beta_i,\lambda_i)$ is unconstrained. To see this, consider fixing the coupling to zero. Then, it does not matter what value we pick for the Compton wavelength or its time dependence since we always recover GR. Similarly, if the Compton wavelength is unobservably small, we expect no constraint on the coupling or time
dependence, since the effects of the scalar field are never felt at observable scales. The posterior is then very non-Gaussian in these parameters around the GR model and an error analysis based on the Fisher matrix will be wrong.
A better approach would be to use some related parametrization which uses specific values of $\mu$ and $\gamma$ at predefined scales and times, which should have a much more Gaussian posterior. One could then map the Fisher constraints onto non-Gaussian distributions for the 
$(\beta_i,\lambda_i)$ parameters.

\section{Intensity mapping survey assumptions}
\label{sec:IM}

In this section we provide details of the window functions and survey parameters
of our assumed 21\,cm experiment, and the astrophysical bias. Throughout,
we use a Gaussian frequency-space window
function normalized to unity with a fixed bandwidth of 2\,MHz. We consider the
frequency range 400--800\,MHz corresponding to $z=0.7$--$2.5$. 
The corresponding radial resolution is $12\,\text{Mpc}$ at $z=0.7$ and
$34\,\text{Mpc}$ at $z=2.5$. Smaller bandwidths are possible and may improve
our constraints somewhat since at the upper end of the redshift range we include
linear modes short enough to be suppressed by integration through the window
(see Sec.~\ref{sec:methods}). However, using a narrower window requires
finer sampling and increases the run-time of the code
significantly.



We take a scale-independent bias of $b=2$, which is consistent with
the bias of DLAs recently derived from cross-correlation with the
Ly$\alpha$ forest~\citep{2012arXiv1209.4596F}. We further take the bias
to be independent of redshift. In our forecasts, we later treat the bias
as a free parameter and marginalise over it, arguing that its exact
numerical value is not important.

For our experimental setup, we consider a small-scale interferometer
consisting of an array of dipole antennae. We take the dimensionful
thermal noise angular power spectrum to be white with~\citep{2004ApJ...608..622Z}
\begin{equation}
\frac{l(l+1)C_l^N}{2\pi} = \frac{T_{\mathrm{sys}}^2(2\pi)^2}{\Delta \nu t_o
  f^2_{\mathrm{cover}}} \frac{l(l+1)}{l_{\mathrm{max}}^2},
\label{eq:noise}
\end{equation}
and the noise uncorrelated between frequency bins. Here, 
$T_{\mathrm{sys}}$ is the system temperature of the antennae,
$\Delta \nu$ is the bandwidth, $t_o$ is the total observing time,
$f_{\mathrm{cover}}$ is the ratio of the total collecting area to the 
geometric area of the array, and $l_{\mathrm{max}}$ is the
maximum multipole given by $2\pi D / \lambda$ where $\lambda$ is the
observing wavelength and $D$ is the diameter of the
array. The quantity $f_{\mathrm{cover}}$ is sometimes called the array filling
factor in the literature, and is closely related to the aperture
efficiency of a single-dish experiment. We set $f_{\mathrm{cover}}=1$
for simplicity, since its precise value depends on the geometry of the
experiment. The system
temperature is given by $T_{\mathrm{sys}} = T_{\mathrm{ant}} + T_{\mathrm{sky}}$
where $T_{\mathrm{ant}}$  is the antenna temperature and
$T_{\mathrm{sky}}$ is the angle-averaged sky temperature due to
foreground contamination. At the low redshifts considered here, the
system temperature is dominated by the antenna temperature, which we
take to be 40\,K, a reasonable estimate given current technology
\citep{2010ARA&A..48..127M}. For the sky temperature we take
\begin{equation}
T_{\mathrm{sky}} = 5.0 \left( \frac{\nu}{710\,\mathrm{MHz}}
\right)^{-2.6} \mathrm{K}
\label{eq:skytemp}
\end{equation}
appropriate for synchrotron radiation.
We note that Eq.~\eqref{eq:noise} has been derived under the
assumption that the array uniformly covers a region of Fourier space of area $\pi
l_{\mathrm{max}}^2$ as the Earth rotates; for refinements accounting for
non-uniformity in the array see e.g.~\cite{2004ApJ...608..622Z}.

We assume a total integration time of one year, and assume the array
has a diameter of $D=100\,\text{m}$, corresponding to something roughly the size
of CHIME~\cite{CHIME}.
In addition, we assume the survey covers a fraction of the sky $f_{\text{sky}} = 0.5$.
Our experimental parameters are summarized in Table~\ref{table:experiment}.
\begin{table}
  \caption{Experimental parameters adopted for our forecasts; see text for definitions.}
  \begin{center}
    \begin{tabular}{cc}
      \hline \hline
      Parameter & Value \\
      \hline
      $\nu_{\text{max}}$ & 835\,MHz \\
      $\nu_{\text{min}}$ & 406\,MHz \\
      $\Delta \nu$ & 2\,MHz \\
      $D$ & 100\,m \\
      $f_{\mathrm{cover}}$ & 1.0 \\
      $t_o$ & 1\,yr \\
      $T_{\text{ant}}$ & 40\,K \\
      $f_{\text{sky}}$ & 0.5 \\
      \hline \hline
    \end{tabular}
  \end{center}
  \label{table:experiment}
\end{table}

Since the sources of the 21\,cm signal are discrete, the measured auto-spectra
have shot noise contributions as well as the clustering signal. For a redshift
window with an angular density of sources $\bar{N}(z)$, we include a dimensionless shot noise
power given by $C_l^{\text{shot}} = 1/\bar{N}(z)$. We assume a comoving number density of sources of $0.03\, h^3 \text{Mpc}^{-3}$ following~\cite{2010PhRvD..81f2001M}.

\begin{figure}
\centering
\includegraphics[width=0.6\textwidth]{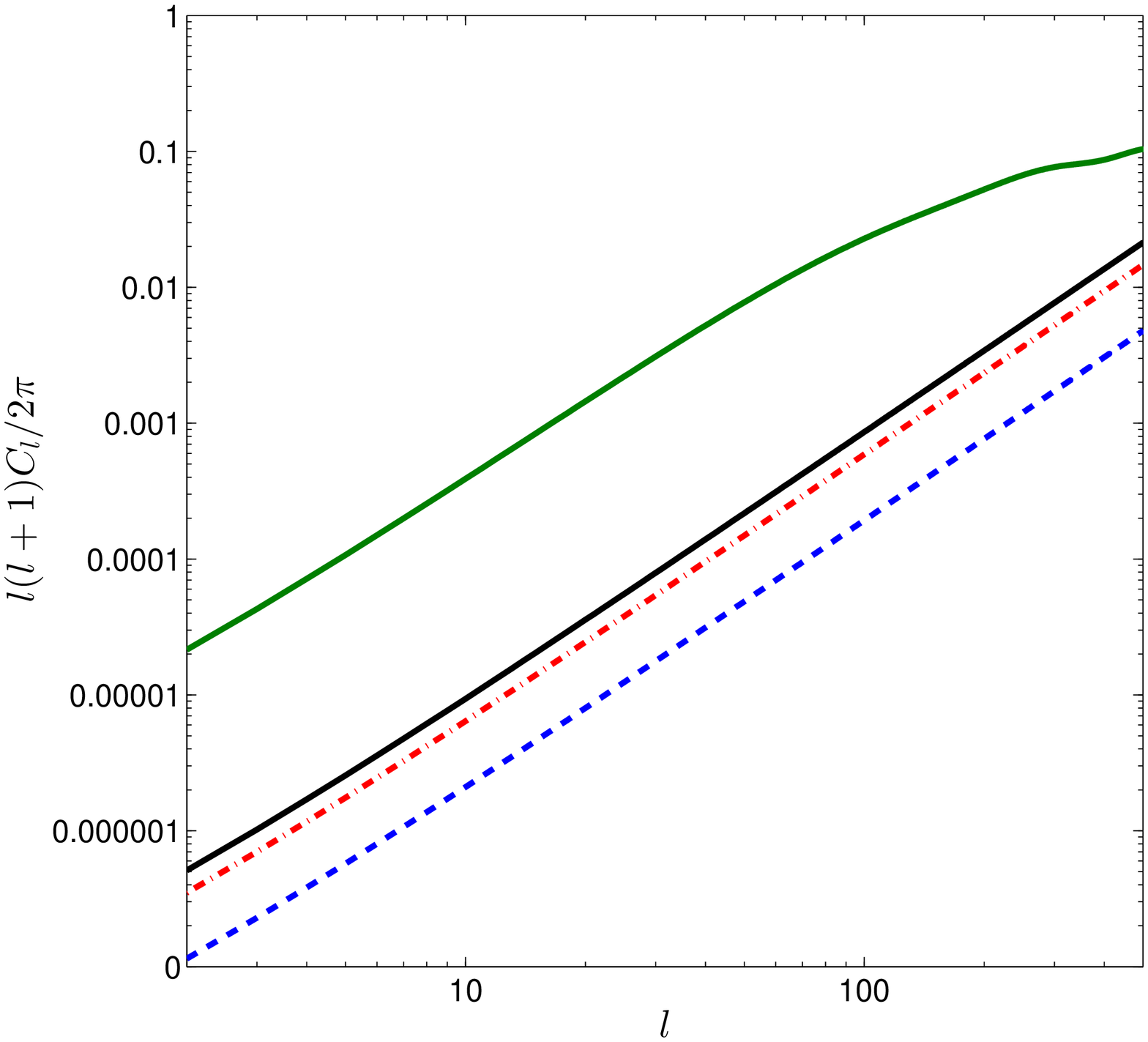}
\includegraphics[width=0.6\textwidth]{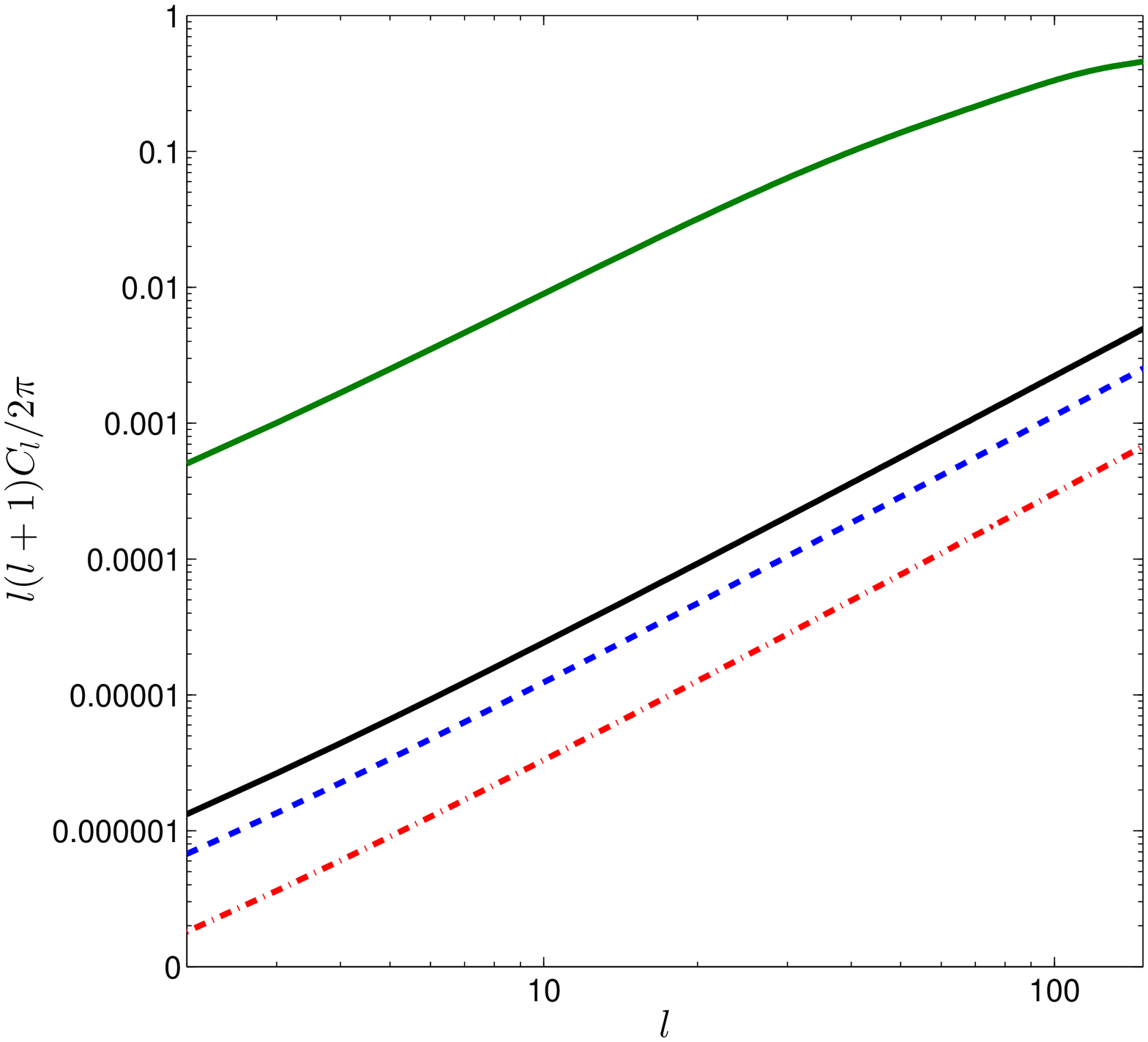}
\caption{(Color Online). Dimensionless 21\,cm brightness temperature auto-spectra (green, upper solid), shot noise power (blue, dashed), thermal noise power (red, dot-dashed) and total noise power (black, lower solid) at $z=2.5$ (top) and $z=1.0$ (bottom), for the survey parameters in Table~\ref{table:experiment}.}
\label{fig:noisespec}
\end{figure}

We further assume that no modes are used with multipole
greater than some $z$-dependent cut-off $l_{\mathrm{max}}^{21\,\text{cm}}(z)$ to avoid
issues with modelling non-linear scales (see Sec.~\ref{sec:methods}).
In Fig.~\ref{fig:noisespec} we plot the dimensionless 21\,cm auto-spectra and noise power spectra at $z=2.5$ and $z=1.0$. In each case, the greatest
multipole plotted is the cut-off $l_{\mathrm{max}}^{21\,\text{cm}}(z)$.

We further split the frequency range into $N_{\text{win}}$ 2\,MHz frequency
windows. The choice of $N_{\text{win}}$ and the spacing of the windows is a subtle issue dependent to some extent on what the experiment is trying to measure. Ideally, we should include all $(\nu_{\text{max}}-\nu_{\text{min}})/\Delta \nu$ windows to retain all information. However, the number of power spectra increases with the number of windows as $N_{\text{win}}^2$ and calculating all of these is computationally prohibitive for the forecasts in this paper. This problem would be exacerbated in a full parameter analysis where the spectra would have to be computed $O(10^5)$ times across the parameter space. Of course, the observed signal will be highly correlated for sufficiently small radial separations. Fine sampling on radial scales smaller than this correlation length will not improve parameter constraints significantly in models with power spectra that vary slowly in time if the measurements in individual windows are sample-variance limited. Note that for 21\,cm cosmology, on large angular scales the radial correlation length is rather smaller than $\chi(z)/l$ since shorter modes contribute significantly there. This means that even if one is interested in a parameter that only affects large scales, it is advantageous to space windows more closely than this to beat down the cosmic variance of the smaller-scale modes. An alternative approach is to transform to a different radial basis to remove the correlations. For example, if we were simply observing projections of a time-independent scalar field, a spherical-Bessel transform in the radial direction (adopting a fiducial cosmology to convert redshift to distance) would yield approximately uncorrelated spherical multipoles. Such an approach is advocated in~\cite{2003MNRAS.343.1327H} for large-scale cosmic shear surveys. Note that galaxy redshift surveys circumvent the issue on small scales by breaking the survey volume up into sub-volumes and estimating the (anisotropic) 3D power spectrum independently in each volume.

A full exploration of these issues is beyond the scope of this paper. Instead, for most of our forecasts we simply take $N_{\text{win}} = 20$ windows equally spaced across the observable frequency range. As we discuss in
Sec.~\ref{subsec:fR_results} and~\ref{sec:PCA}, we expect our qualitative results to be stable to increases in $N_{\text{win}}$ and to changes in their location. However, in our PCA of general models we do expect that the constraints on the amplitudes of individual modes could be improved though not their general form or the number of well-determined modes.


\section{Forecasting methodology and results}
\label{sec:methods}

In this section we present the details of our statistical analysis and forecasted constraints on the parameters of modified gravity.

We construct a Fisher matrix for the parameters $\theta_i$
(seven cosmological + $N$ modified gravity), for an upcoming 21\,cm intensity mapping experiment with noise spectrum given by Eq.~\eqref{eq:noise} and survey parameters from Table~\ref{table:experiment}. In addition we include CMB (unlensed) temperature, $E$-mode polarization and a reconstruction of the lensing deflection
field $d$ from the CMB. We adopt Planck-like noise levels and and angular resolutions taken from~\cite{2006astro.ph..4069T}, and retain
multipoles up to  $l_{\mathrm{max}}^{\mathrm{CMB}}=2000$. Statistical noise on the lensing deflection reconstruction is computed using the optimal quadratic estimator of~\citep{okamotoandhu}.

The Fisher matrix for the parameters $\theta_i$ is defined as the ensemble average of the Hessian matrix of the log-likelihood, and, assuming Gaussian fields with zero mean and covariance $\mathbf{C}$, is given by 
\begin{eqnarray}
F_{ij} &=& \frac{1}{2}\mathrm{Tr}\left[\mathbf{C}^{-1}
\frac{\partial \mathbf{C}}{\partial \theta_i}
\mathbf{C}^{-1}
\frac{\partial \mathbf{C}}{\partial \theta_j} \right] \nonumber \\
&=& \sum_{l=2}^{l_{\mathrm{max}}}\sum_{XX',YY'}\frac{\partial C_l^{XX'}}{\partial \theta_i}[\mbox{Cov}(XX',YY')]^{-1}_l\frac{\partial C_l^{YY'}}{\partial \theta_j}, 
\label{eq:fisher_matrix}
\end{eqnarray}
where $\mbox{Cov}(XX',YY')$ is the covariance of the power spectra estimators, including noise, and $XX'$ and $YY'$ stand for combinations of the observed fields ($TT$, $TE$, $Ed$ etc.), see for example~\citep{1997ApJ...480...22T}. For the specific form of $\mbox{Cov}(XX',YY')$, see for example~\citep{2012MNRAS.425.1170H}. Note that we include all the cross-correlations between the CMB and 21\,cm fields, which potentially carry a lot of information about late-time phenomena (see \citep{2007PhRvD..76h3518S} for a discussion in the context of galaxy surveys). The non-zero cross-correlation between the low redshift 21\,cm signal and the (unlensed) CMB temperature anisotropy is due to the late-time ISW effect in the CMB\footnote{Note that the 21\,cm signal also contains an ISW effect [see Eq.~\eqref{eq:10}] that is perfectly correlated with the ISW contribution to the CMB for $z>2$.
However the amplitude of this correlation is subdominant with respect to the 21\,cm density-ISW cross-correlation.}, whilst the 21\,cm-polarization correlation arises since the last-scattering surface of an electron at reionization has
some overlap with the radial distances of relevance for the 21\,cm signal~\citep{2011JCAP...03..018L}.  The latter correlation is small, but grows with redshift and we include it for consistency. The inverse of the Fisher matrix gives the covariance matrix between the parameters and its diagonal elements give the $1\sigma$ marginalised errors on parameters.

We vary seven cosmological parameters, $(\Omega_bh^2, \Omega_ch^2, h, \tau, A_s, n_s, \Omega_{\nu}h^2)$, and assume three degenerate neutrinos at the minimum mass of the normal hierarchy ($\Omega_{\nu}h^2 \approx 0.0006$). We fix the curvature to be zero, and the expansion history to be that of $\Lambda$CDM (plus massive neutrinos),
as favoured by a variety of cosmological probes \citep{2011ApJS..192...18K}.



To avoid the issues of modelling non-linear structure formation in modified gravity,
for the 21\,cm signal we retain only multipoles up to $l_{\mathrm{max}}^{\mathrm{21\,cm}}$. Whilst the comoving non-linear scale at the present epoch in GR is roughly at $k_{\mathrm{NL}} \approx 0.1 \, h \mathrm{Mpc}^{-1}$, in modified gravity it can be rather larger. For example, $N$-body simulations of $f(R)$ models in \citep{2012MNRAS.425.2128J} show that linear physics is only recovered for $k < 0.06 \, h \mathrm{Mpc}^{-1}$ at $z=0$. To account for these effects, we set $l_{\mathrm{max}}^{\mathrm{21\,cm}} = 150$ for $z < 1.5$, corresponding to $k_{\mathrm{NL}}^{\mathrm{21\,cm}} \approx 0.05$--$0.08 \, h \mathrm{Mpc}^{-1}$, which roughly matches the non-linear scales found in \citep{2012MNRAS.425.2128J}\footnote{Note that we fix the maximum $l$ rather than the maximum $k$.}. For $z>1.5$, we set $l_{\mathrm{max}}^{\mathrm{21\,cm}} = 500$, corresponding to $ k_{\mathrm{NL}}^{\mathrm{21\,cm}} \approx 0.12$--$0.15 \, h \mathrm{Mpc}^{-1}$, since the comoving non-linear scale is still expected to grow with time in modified gravity models.
We ignore non-linearities in the CMB lensing deflection power spectrum since,
at Planck noise levels, the signal-to-noise on the lensing reconstruction is
expected to be very low on non-linear scales.

To include the effects of modified gravity in \textsc{CAMB sources}, we use the \textsc{MGCAMB} module \citep{2009PhRvD..79h3513Z,2011JCAP...08..005H}, adapted for compatibility with \textsc{CAMB sources}. We also update the sub-horizon radiation approximations to those of the full \textsc{CAMB} release. The May 2011 release of \textsc{CAMB sources} uses the old sub-horizon approximations of \citep{2005JCAP...06..011D}, but this formalism implicitly assumes GR. We update this part of the code to use the newer formalism of \citep{2011JCAP...07..034B}, which can be easily generalised to include modified gravity. We also incorporate the effects of anisotropic stress from massive neutrinos consistently in \textsc{MGCAMB}, although the effects are negligible at late times.

\subsection{Constraints on $f(R)$ gravity}

\label{subsec:fR_results}

We start by considering the constraints that 21\,cm intensity mapping can place on $f(R)$ theories. In this section, we fix the bias to a constant value of $b=2.0$, consistent with the constraints from \cite{2012arXiv1209.4596F}.
Note that to satisfy local tests of gravity, the scale of the $f(R)$ Compton wavelength $\lambda$ at present cosmological density is restricted to less than around
1\,Mpc~\citep{2012arXiv1208.4612W, 2012PhLB..715...38B}. This limits departures from GR in $f(R)$ theories consistent with Solar System tests to non-linear scales. However, here we are not interested in constraining $f(R)$ theories per se, but rather in understanding which kind of constraints 21\,cm observations could place on an (effective) theory that behaves like $f(R)$ gravity at cosmological scales, without necessarily satisfying local tests.

We consider the $B_0$-parametrization of $f(R)$ gravity~\citep{2007PhRvD..76f4004H}, which provides a good approximation on quasi-static scales \citep{2010JCAP...04..030G,2012PhRvD..86l3503H}. This is closely related to the scalar-tensor type parametrization discussed in Sec.~\ref{sec:MG}, since $f(R)$ is conformally equivalent to a scalar-tensor theory. The specific forms of $\mu$ and $\gamma$ we consider are
\begin{eqnarray}
\mu(a,k) &=& \frac{1}{1 - 1.4 \times 10^{-8} \lvert \lambda/ \text{Mpc} \rvert^2 a^3} \frac{1+4\lambda^2k^2a^4/3}{1+\lambda^2k^2a^4},
\label{eq:fRmu} \\
\gamma(a,k) &=& \frac{1+2\lambda^2k^2a^4/3}{1+4\lambda^2k^2a^4/3},
\label{eq:fRgamma}
\end{eqnarray}
where the parameter $B_0 \equiv 2 H_0^2 \lambda^2$ is the square of the
Compton wavelength of the effective scalar degree of freedom $f_R \equiv \ud f / \ud R$ (the
\emph{scalaron}) in units of the current Hubble length.
The prefactor in Eq.~\eqref{eq:fRmu} accounts for the time-dependence of Newton's constant in the Jordan-frame background equations, but is practically unity for the models we consider. It is then easy to show that $\Sigma \approx 1$ in this theory, so the relationship of the Weyl potential to the density is the same as in GR (although the density is of course different).

Our parameter space thus consists of the seven cosmological parameters, plus $B_0$. We take GR as a fiducial model, i.e. $B_0 = 0$ and adopt 20 window functions equally spaced in frequency.
Our $1\sigma$ errors are $B_0$ are $1.9 \times 10^{-4}$ using only the CMB information, improving to
$3.7 \times 10^{-5}$ with the addition of 21\,cm. When the analysis is run without any CMB information, the $1\sigma$ constraint on $B_0$ is $6.9 \times 10^{-5}$. This suggests that the constraint is driven primarily by the 21\,cm observables. Put another way, adding 21\,cm information improves the constraint by a factor of five over the CMB alone, whereas adding CMB information improves the constraint by a factor of two over 21\,cm alone.

We have marginalised over all other parameters, but there is only mild degradation in the constraint as a result of this. The main covariance of $B_0$ is with $\Omega_{\nu}h^2$ (correlation coefficient of $0.4$). Increasing $B_0$ pushes the scalaron Compton wavelength to larger scales, such that a greater range of (small) linear scales experience enhanced growth and for longer. This boosts the 21\,cm signal. The main effect of light massive neutrinos on the 21\,cm power spectrum is a suppression of power on scales below the horizon size at the redshift where the neutrinos become non-relativistic. The suppression is scale-dependent for scales between the free-streaming scale at the transition and the free-streaming scale at the 21\,cm redshift, with smaller scales exceeding the free-streaming length later and hence suppressing the growth of structure for longer~\citep{2012MNRAS.425.1170H}. On scales smaller than the free-streaming length at the 21\,cm redshift the suppression is scale-free, and for scales larger than the free-streaming length at the non-relativistic transition there is no suppression at all. This small-scale suppression can partially cancel the enhancement from increasing $B_0$, giving a positive correlation.


%
%
As noted above, the addition of information from 21\,cm intensity mapping improves the constraint on $B_0$ by about a factor of five over the CMB-alone case.  Although this is not competitive with local constraints or forecasted constraints for the Dark Energy Survey or Large Synoptic Survey Telescope \citep{2012PhRvD..85d3508H}, it demonstrates the potential usefulness of intensity mapping in constraining $f(R)$ theories. It is interesting to note that our constraints are comparable with those in~\cite{2010PhRvD..81f2001M}, who considered only the baryon acoustic oscillation signature
extracted from 21\,cm maps, lens reconstruction from 21\,cm itself and Planck-like
priors, when only linear scales are included.
To map from their parametrization to ours, note that $B_0 \approx
- 2(n+1) f_{R_0}/(4-3\Omega_m)$ where the power-law $n$ appears in the
Hu \& Sawicki model of $f(R)$~\cite{2007PhRvD..76f4004H} that they employ. Current cosmological constraints from \cite{PhysRevD.85.124038} limit $B_0 < 1.1\times 10^{-3}$ at 95\% confidence, although this constraint is driven mainly by cluster abundance data where non-linear chameleon effects not accounted for in \cite{PhysRevD.85.124038} could be important. Excluding the cluster abundance data, \cite{PhysRevD.85.124038} quote an upper limit of $B_0 < 0.42$ at 95\% confidence. Our 95\% upper limit is $7.4\times 10^{-5}$, which represents an improvement on this current constraint by roughly four orders of magnitude.

One might question why our CMB-alone constraint on $B_0$ improves so impressively over current CMB constraints from WMAP~\cite{2007PhRvD..76f3517S}.
Almost all of the improvement can be attributed to the information brought by CMB lensing, not present in the WMAP constraint. We note that one would not expect a simple Fisher analysis to reproduce accurately measured constraints on $B_0$ from WMAP, since this constraint comes entirely from large angular scales (the
late-time ISW effect) where one has to worry about non-Gaussianity of the likelihood as well as the effects of finite sky coverage.

What limits the power of intensity mapping in constraining $B_0$? The main effect of $B_0$ on the 21\,cm power spectrum is an enhancement of power on scales below the Compton wavelength, with the enhancement increasing with time. Thus, setting aside the influence of the CMB for now, we may conjecture that the upper limit set on $B_0$ from 21\,cm data derives from the shortest observable scale at the lowest observable redshift. In our case, this is the non-linear scale\footnote{Note that our choice for the non-linear scale is more conservative than in~\citep{2012PhRvD..85d3508H}. This may partly explain their stronger constraint on $B_0$.}, which at $z=0.7$ is roughly $0.08 \, h \mathrm{Mpc}^{-1}$. The value of $B_0$ which sets the comoving Compton wavelength equal to this scale at $z=0.7$ is around $3 \times 10^{-4}$. This is only a rough upper limit, since $\mu$ and $\gamma$ are not step functions, and even scales above the Compton wavelength experience some modified growth.
We can therefore anticipate that as our understanding of modified-gravity
scenarios in the mildly non-linear regime of structure formation improves, we will
be able to exploit 21\,cm (and other cosmological observables) on smaller scales
thus enhancing our ability to constrain scalar-tensor theories and $f(R)$.

With some experimentation, we find that our results are stable to changes in the number of windows around $N_{\text{win}} = 20$, equally spaced across the observable frequency range. For the $B_0$ constraints, most of the signal comes from the lowest observed redshifts, so adding more bins does not impact the constraints significantly. Our results are also stable to moving all 20 bins to low redshift, with a radial separation corresponding to the distance at which the cross-correlation with the $z=0.7$ window at $l=150$ first goes to zero. 

To understand which part of the 21\,cm signal is most powerful in constraining $B_0$, we repeat the analysis with the density term switched off, in both cosmic variance and the signal. This is equivalent to what could be achieved if the redshift-space
distortion signal could be extracted coherently in the 21\,cm maps. While
not possible for 21\,cm intensity mapping, we note that such extraction is
possible if the clustering of differently biased populations is combined~\cite{2009JCAP...10..007M}.
We find that the marginalised error on $B_0$ \emph{decreases} by roughly 60\%, suggesting that redshift-space distortions
are significantly more sensitive to $B_0$ than the density,
and that the information we can extract from redshift-space
distortions is limited by the cosmic variance of the dominant density term.

On the CMB side, one question we can answer quite easily is whether it is CMB lensing or the ISW which is more powerful in constraining $B_0$ in our combination of observables. To investigate this, we run the above analysis again but with the late ISW term set to zero. We find that the marginalised error on $B_0$ increases by roughly 0.5\%, indicating that the information about modified gravity brought by the late ISW effect is sub-dominant to that brought by CMB lensing. This is likely due to large cosmic variance present in the CMB at the large angular scales where the late-time ISW effect is significant.


\subsection{Principal component analysis}
\label{sec:PCA}
Whilst specific parametrizations of the modified gravity sector such as Eqs.~\eqref{eq:BZmu} and \eqref{eq:BZgamma} have the advantage of being physically transparent and easily related to the underlying theory, they are too restrictive in scope and fail to describe the general space of theories. To test GR
on cosmological scales in a more theory-agnostic manner, a model-independent
approach is required whereby one considers rather general functional forms for $\mu$ and $\gamma$. Such an approach is provided by the machinery of PCA.
The basic idea is to establish what functional forms of $\mu(a,k)$ and
$\gamma(a,k)$ can be independently constrained well by the data.

The first step in a PCA is the pixelization of the free functions, ideally at fine enough
resolution that the well-determined modes coincide with what would be obtained from
a continuum analysis. In practice, for this forecasting analysis, computational
resources limit us to consider $4 \times 4$ grids for both $\mu$ and $\gamma$, equally spaced in both comoving wavenumber and conformal time. Our range of $k$ is $0.00 < k <  0.153 \, h\mathrm{Mpc}^{-1}$, and our redshift range is $0 < z < 3$. The upper limit on $z$ is arbitrary, but we only wish to consider models where modified gravity mimics dark energy, which motivates a choice of upper limit that matches the onset of acceleration. We enforce GR for $z>3$. The upper limit on $k$ corresponds to the smallest observable scale in the 21\,cm signal.
For $k>0.153 \, h\mathrm{Mpc}^{-1}$ we set our functions equal to their values at $k=0.153 \, h\mathrm{Mpc}^{-1}$. The choice of grid resolution imposes a prior
on how rapidly we allow the modified-gravity functions to vary in scale and time.

We thus parametrize departures of $\mu$ and $\gamma$ from their GR form (unity)
with 16 nodal values each.
We use a bicubic spline to interpolate between these nodes, setting the derivatives to zero at the edges of the grid. We also experimented with bins rather than nodes, in the manner of \citep{2009PhRvL.103x1301Z,2012PhRvD..85d3508H}, but found that the steep variation of $\mu$ and $\gamma$ at the bin edges caused a loss in numerical accuracy. We transform our Fisher matrix into the $(\mu,\Sigma)$ basis, since conditional information from $\Sigma$ comes almost completely from the CMB alone,
whereas information about $\mu$ comes from both 21\,cm and CMB fields.

\subsubsection{Fixed bias, $b=2$}
\label{subsec:fixedb}

\begin{figure}
\centering
\includegraphics[width=0.5\textwidth]{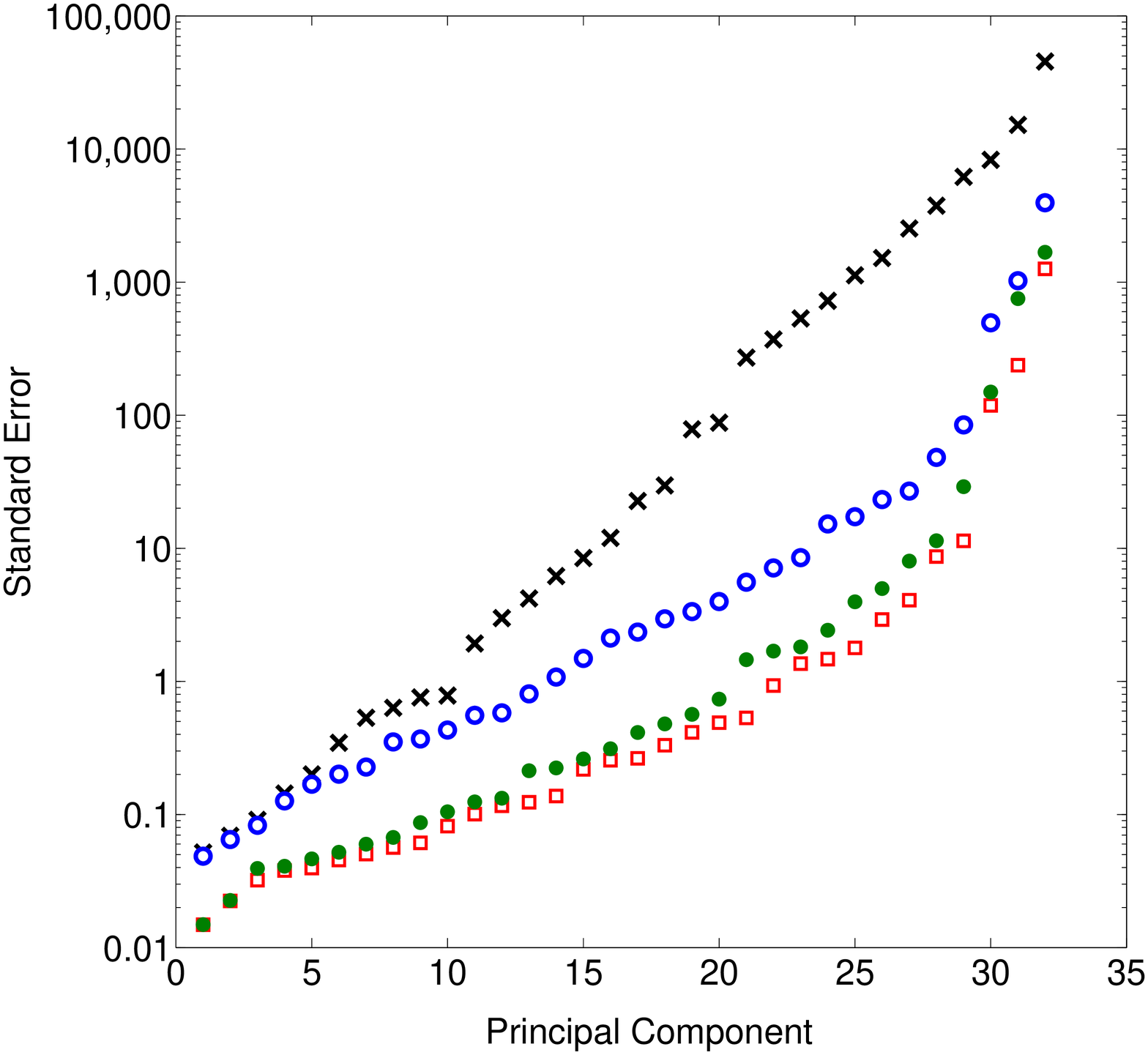}
\caption{(Color Online). The $1\sigma$ errors on the amplitude of each mode of the PCA (i.e.\ the square root of each eigenvalue) from the CMB alone (black crosses) and 
CMB with 21\,cm, throwing away information from the density term (blue open circles), marginalising over 20 scale-independent bias parameters (green filled circles) and with a fixed constant bias (red open squares).}
\label{fig:eigvals_full}
\end{figure}

We first report results of the PCA assuming the bias $b=2$ is known perfectly.
In this case, our modified-gravity sector consists of 32 parameters and our total parameter space has a dimension of 39. We construct the Fisher matrix for all
these parameters, and marginalise over the seven cosmological parameters by removing the appropriate rows and columns from the full covariance matrix. We then proceed to diagonalise this sub-matrix, which gives us 32 eigenvectors and associated eigenvalues. Each eigenvector is normalized to unity, and gives an independently constrained mode of $\mu$ and $\gamma$ for our observables. The $1\sigma$ marginalised errors in the determination of the amplitude of each of these
normalised modes is given by the square-root of the corresponding eigenvalue. We plot these errors in ascending order in Fig.~\ref{fig:eigvals_full}.

Since the eigenmodes of the PCA refer to departures from GR, and so have
fiducial value zero, there is no unambiguous definition of a `well-constrained' set of principal components. However, the observations only provide useful information
on those modes for which a $1\sigma$ fluctuation in amplitude corresponds to
changes in $\mu$ and $\Sigma$ which are not large compared to unity. Where this is not the case, physicality priors, such as $\mu > 0$,
would constrain the amplitude of the mode much more strongly than the data.
Moreover, it is clear that those modes for which a $1\sigma$ fluctuation in amplitude
produce large changes in $\mu$ and $\Sigma$ do not give a precise test of GR.
Since the eigenvectors are normalised to unity in a 32-dimensional space, the
typical eigenvector will have at least one component that is $O(1)$ and so we regard
eigenvalues less than one as corresponding to well-constrained modes\footnote{Note that this conclusion would change if we used many more grid-points. The best
determined modes are reasonably well sampled by our grid (see below) and so
we expect these would not change significantly with greater grid resolution. However,
the components would reduce in amplitude, to preserve the normalisation, and the
eigenvalues increase to preserve the form of $\mu$ and $\Sigma$ for an amplitude
of $1\sigma$. The threshold eigenvalue for well-determined modes would thus
increase with grid resolution. This makes direct comparisons with other work, such
as~\citep{2012PhRvD..85d3508H} which considered future cosmic shear and galaxy surveys, difficult since the dimensionality of the grids differ.}. On this measure,
our combination of observables can constrain well around 22 modes of modifications
to gravity.
The discernible upturn in the spectrum of eigenvalues in Fig.~\ref{fig:eigvals_full}
for the 21\,cm-plus-CMB combination suggests that we have reached the saturation point of the PCA, in that the data would not be informative about the
additional modes that would result from increasing the number of grid points.
The upturn is at an $O(1)$ eigenvalue which makes our enumeration of the
well-determined modes reasonably robust to changes in the threshold eigenvalue.
We see further from Fig.~\ref{fig:eigvals_full} that adding 21\,cm information to the CMB observables increases the number of well-constrained modes considerably, as well as reducing the overall errors on these modes. This is due to the information on scale and time-dependence of $\mu$ brought by the 21\,cm signal, as well as degeneracy breaking effects in the combination of CMB and 21\,cm, which we discuss below.

\begin{figure}
\centering
\includegraphics[width=0.4\textwidth]{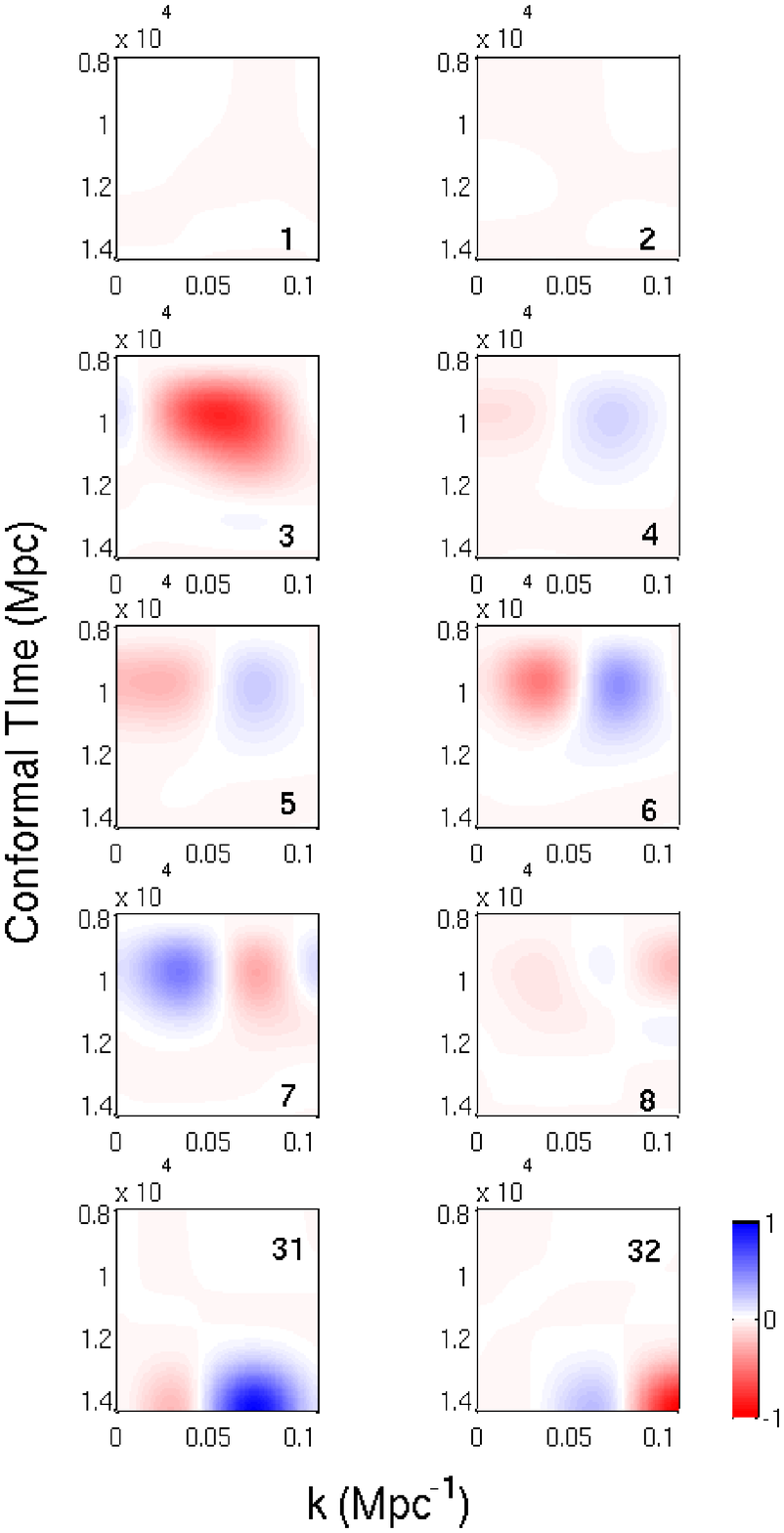}
\includegraphics[width=0.4\textwidth]{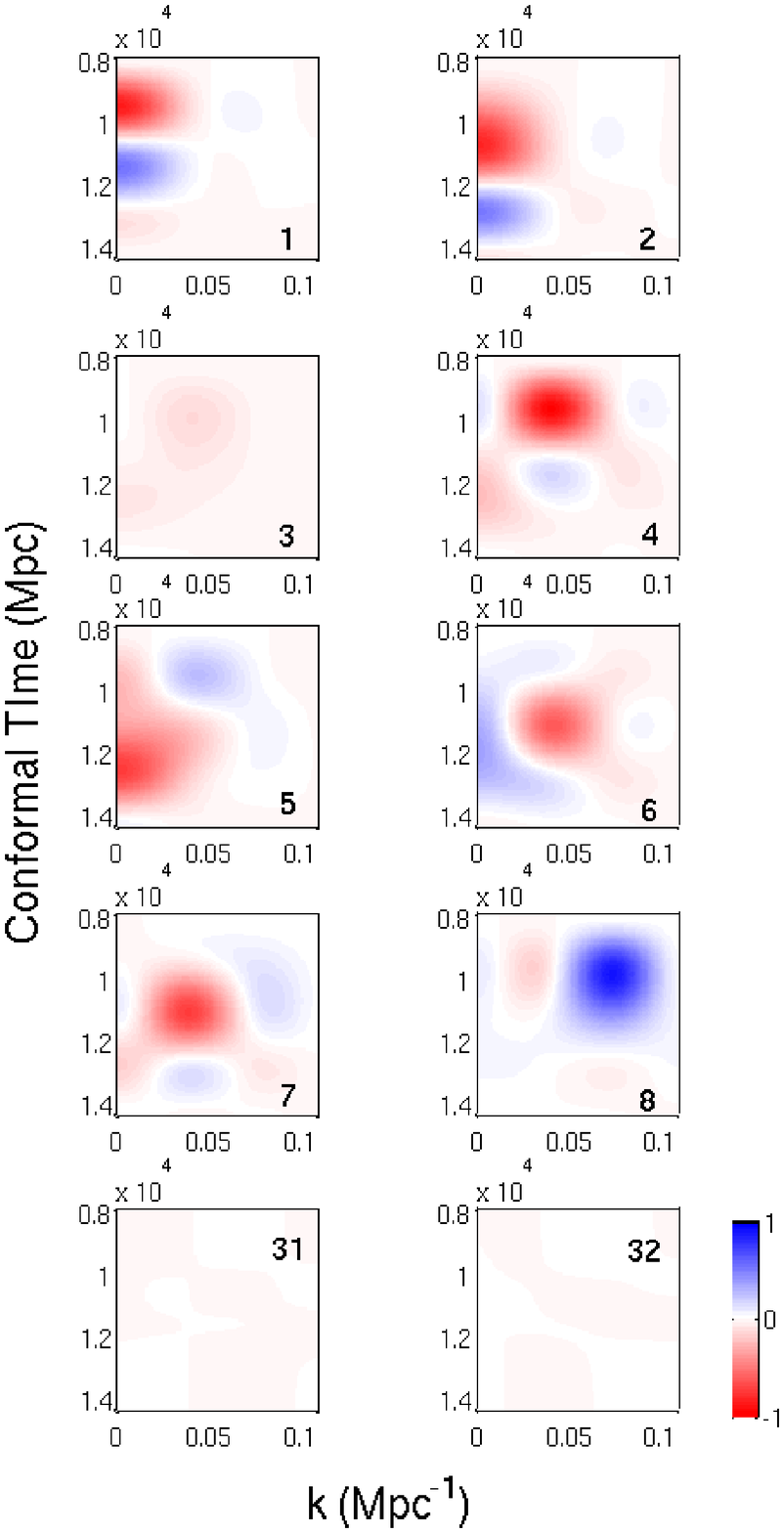}

\caption{(Color Online). \emph{Left}: The $\mu(\eta,k)$ eigensurfaces of the eight most well-constrained and two poorest-constrained
eigenmodes for 21\,cm-plus-CMB observables in the ($\eta,k$) plane. The eigenvectors have been normalised to unity, and the surfaces are bicubic splines in between the nodes of the PCA. \emph{Right}: The same, but for $\Sigma(\eta,k)$.
Note that the overall normalisation is fixed to unity across $\mu$ \emph{and} $\Sigma$ but the sign of each eigenvector is arbitrary.}
\label{fig:eigensurfaces}
\end{figure}

For each eigenvector, we can use its components in the $(\mu,\Sigma)$ basis to construct an `eigensurface' in the $(k,\eta)$ plane for each function, using bicubic splines to interpolate between the nodal values; see
Fig.~\ref{fig:eigensurfaces} for the first eight well-constrained modes
and the two poorest-constrained modes.
Such plots give a visual impression of the `sweet spots' of $\mu$ and $\Sigma$ probed by our particular observables. The poorest-constrained modes
are localised to late times and small scales where our observables have little sensitivity.
Note that we do not plot marginal constraints on either one of $\mu$ or $\Sigma$
alone. We would expect $\mu$ and $\Sigma$ to depend on functions or parameters more fundamental to the underlying theory, as is the case in scalar-tensor theories. It would thus be misleading to reduce the parameter space by marginalising over either of $\mu$ or $\Sigma$ (as well as the cosmological parameters).

We see from Fig.~\ref{fig:eigensurfaces} that the first two eigenmodes are dominated by $\Sigma$, both showing a single broad oscillation in the temporal direction (which
is over-sampled by the grid) and confined to large scales. Although the 21\,cm signal carries no information about $\Sigma$ on sub-horizon scales, the information it brings on $\mu$ is complementary to that of the CMB, and breaks some of the degeneracies between $\mu$ and $\Sigma$ present in the CMB. The source of this degeneracy can be understood roughly as follows: Increasing $\mu$ enhances the density perturbation on sub-horizon scales, which then increases the depth of the Weyl potential via the Poisson equation. This can be compensated by decreasing $\Sigma$, although the cancellation is not exact since the response of the density to $\mu$ is non-local in time. Since the CMB probes modified gravity only through the Weyl potential, there is an approximate degeneracy. Thus, the shapes of the $\Sigma$ eigensurfaces are not the same as those that use only the CMB as an observable. 

\begin{figure}
\centering
\includegraphics[width=0.5\textwidth]{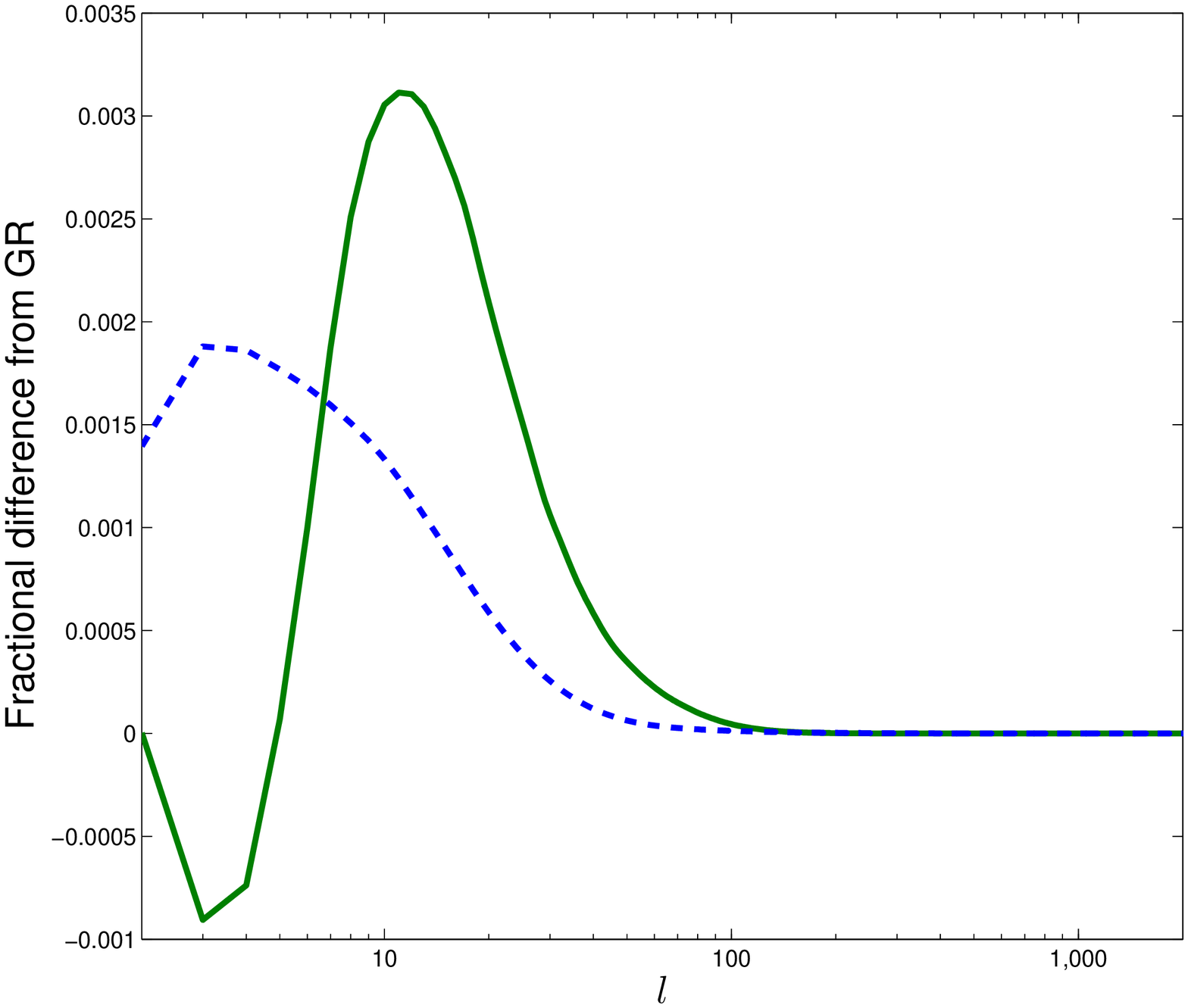}
\\
\includegraphics[width=0.5\textwidth]{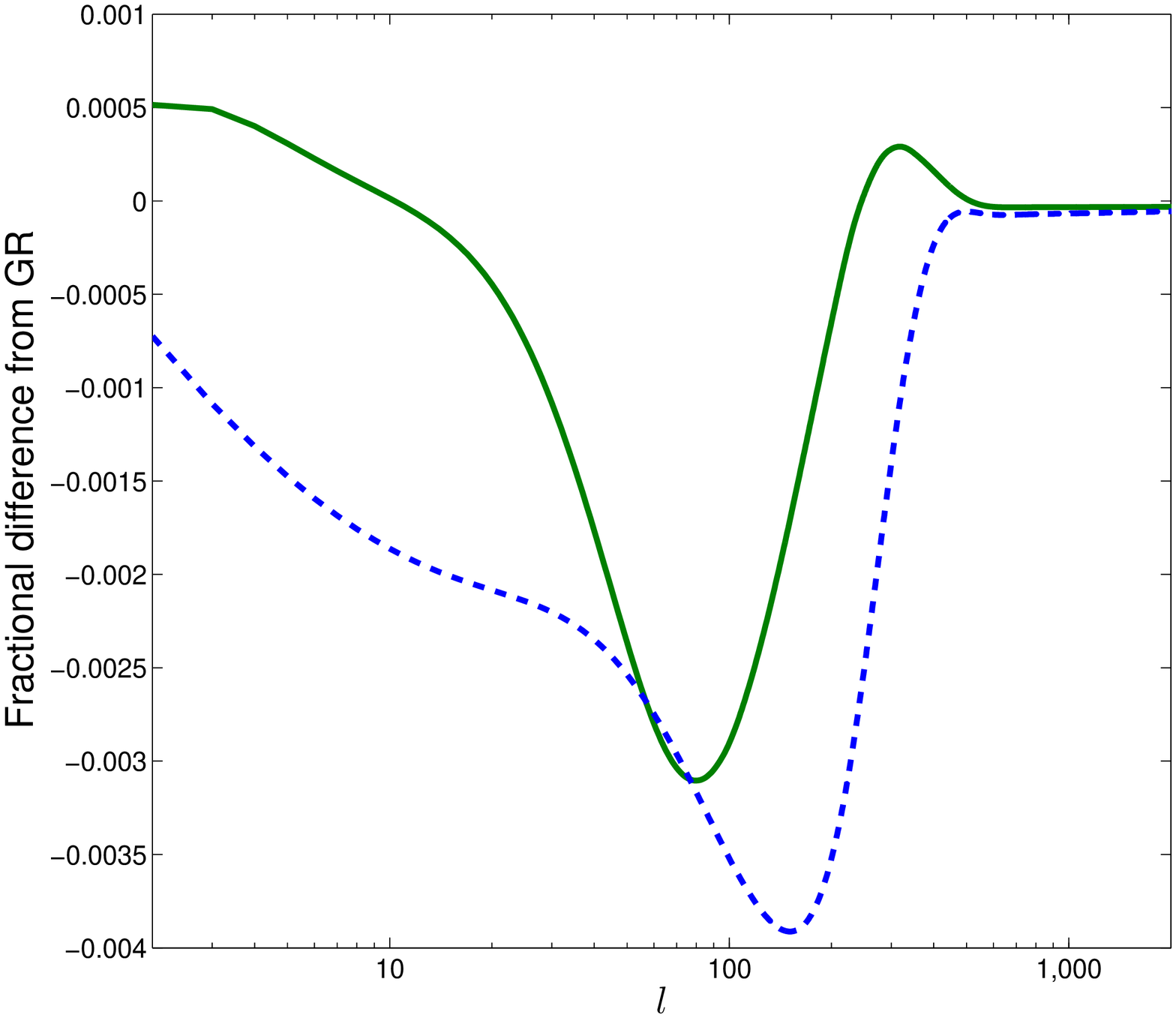}
\caption{(Color Online). Fractional difference from GR in the
CMB temperature power spectrum (top) and
CMB lensing deflection power spectrum (bottom) for $1\sigma$ changes in the
first (green, solid) and third (blue, dashed) best-determined
eigenmodes of the CMB-plus-21\,cm combination.}
\label{fig:TT_dd_13}
\end{figure}

\begin{figure}
\centering
\includegraphics[width=0.5\textwidth]{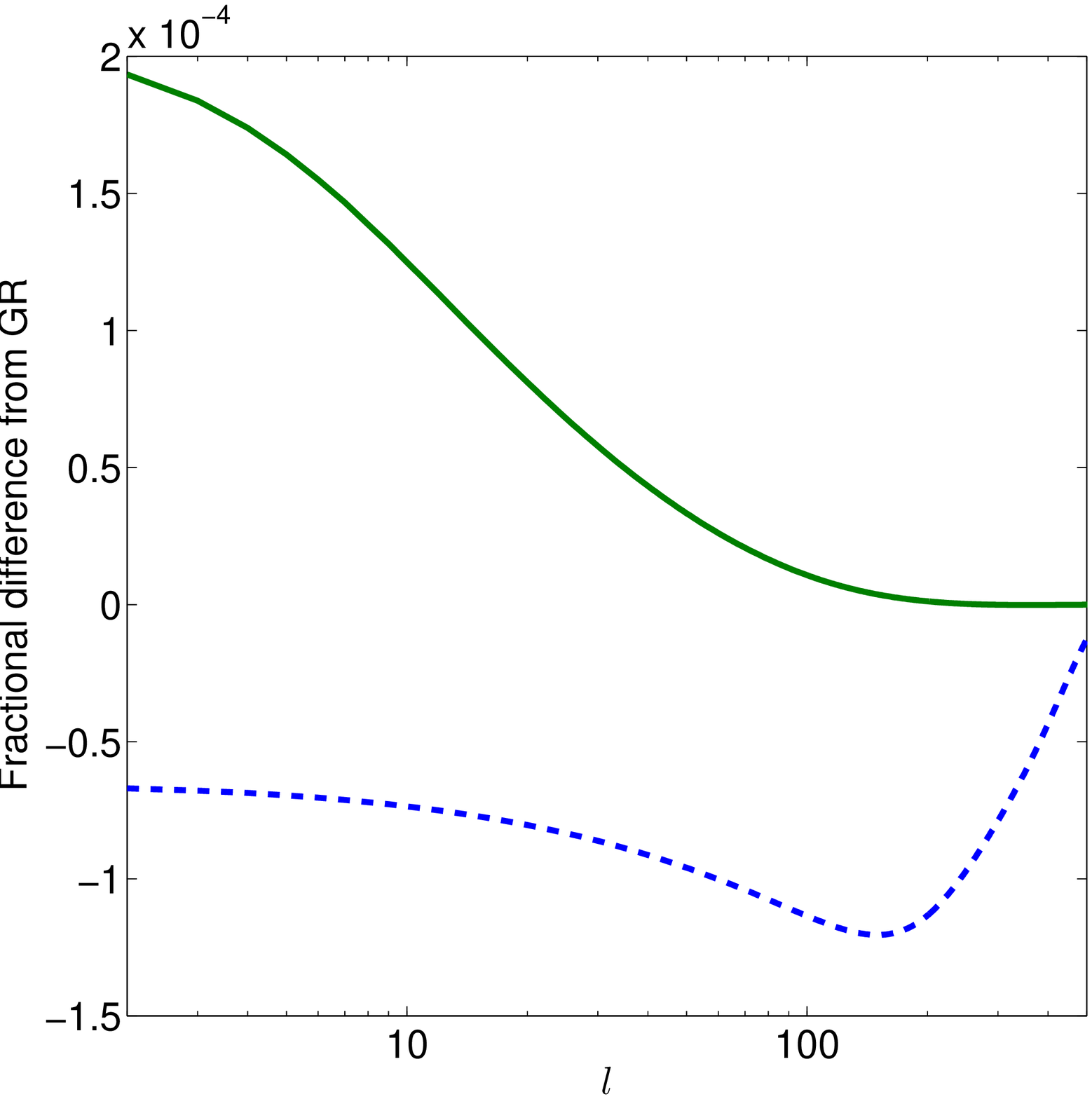}
\\
\includegraphics[width=0.5\textwidth]{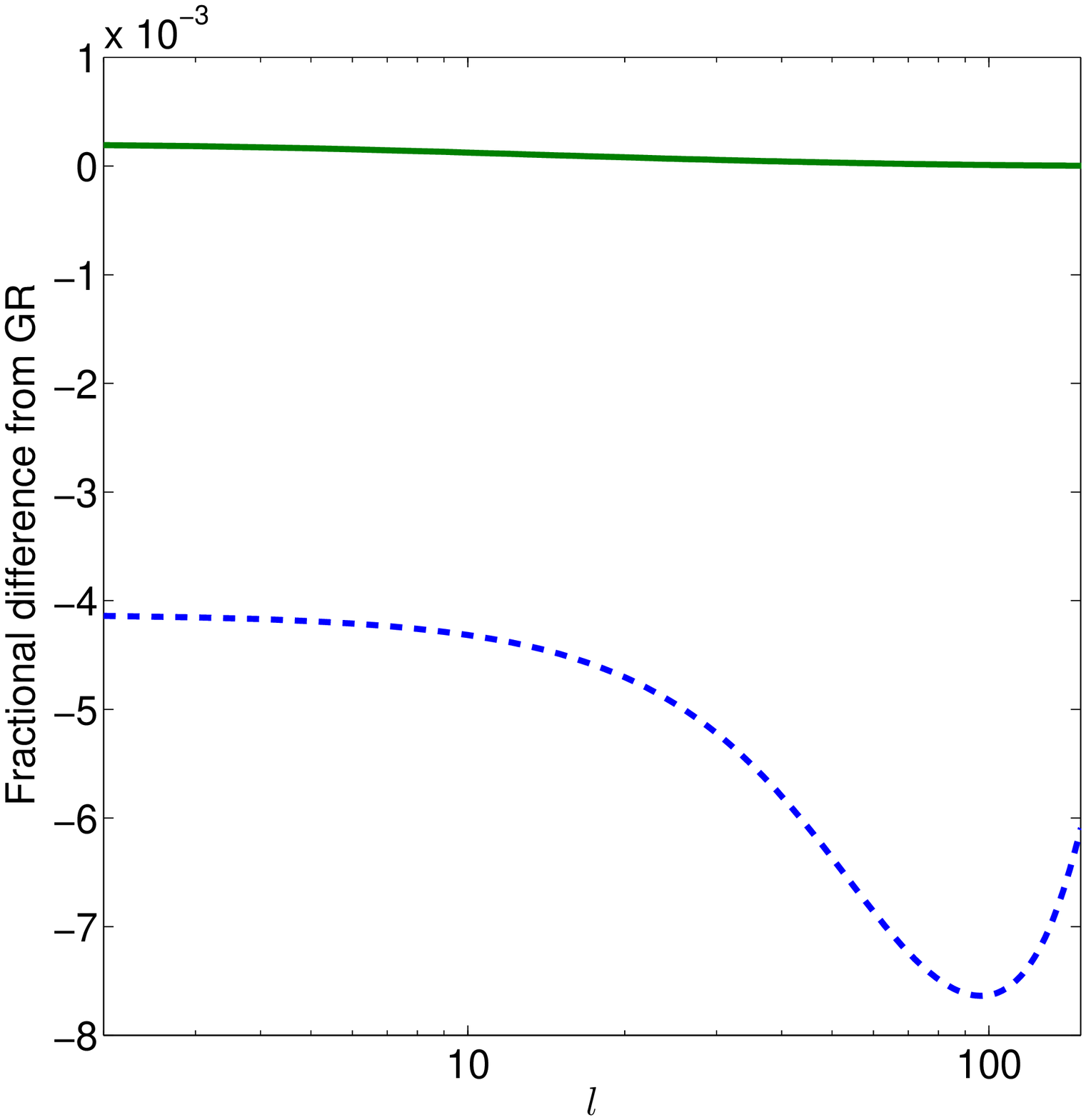}
\caption{(Color Online). As Fig.~\ref{fig:TT_dd_13} but for the 21\,cm power
spectrum at $z=2.5$ (top) and $z=0.7$ (bottom).}
\label{fig:z1z1_z3z3_13}
\end{figure}

The primary sensitivity is thus to temporal oscillations in $\Sigma$, and variations in scale in $\mu$. In Figs~\ref{fig:TT_dd_13} and \ref{fig:z1z1_z3z3_13} we plot the fractional difference of the resulting power spectra from GR for the first and third eigenmodes with amplitude of $1\sigma$. By construction, for both modes plotted the
resulting change in the marginal log-likelihood is $-1/2$.
Note that all differences shown are well within cosmic variance and experimental noise, but the broad-band features allow us to average over many modes and get a constraint.  

We see from Fig.~\ref{fig:TT_dd_13} that temporal oscillations in $\Sigma$ source a large ISW effect in the CMB, since the Weyl potential has a significant time derivative. There is also a lensing contribution, since the oscillations are broad enough that the CMB lensing kernel changes significantly over their period. The information on scale-dependence brought by lensing is of course limited by the Planck reconstruction noise.

\begin{figure}
\centering
\includegraphics[width=0.5\textwidth]{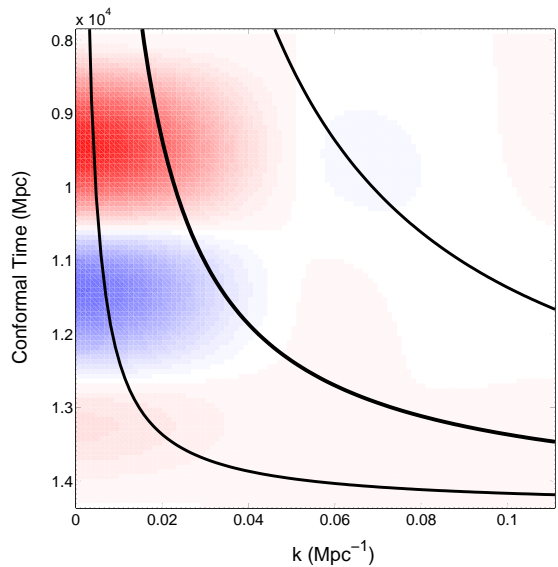}
\caption{(Color Online). The $\Sigma(\eta,k)$ eigensurface of the most well-constrained eigenmode for 21\,cm-plus-CMB observables. The solid black curves correspond to fixed $l=k(\eta_0 - \eta)$ at values of 300 (top), 100 (middle), and 20 (bottom).}
\label{fig:sig_mode1_lensing}
\end{figure}

To understand the shape of the lensing spectrum differences in
Fig.~\ref{fig:TT_dd_13}, consider the power spectrum of the lensing deflection $d$ in the Limber limit, which should be accurate on small angular scales~\citep{2006PhR...429....1L}:
\begin{equation}
C_l^{dd} \approx \frac{8\pi^2}{l}\int^{\chi_*}_0 \ud \chi \, \P_{(\Psi+\Phi)/2}(l/\chi ; \eta_0 - \chi)\frac{(\chi_* - \chi)^2}{\chi_*^2 \chi},
\label{eq:limber}
\end{equation}
where $\P_{(\Psi+\Phi)/2}$ is the dimensionless power spectrum of the Weyl potential, and $\chi_*$ is the conformal distance to recombination. The integral is taken along hyperbolae of constant $l=k(\eta_0 - \eta)$ in the $(\eta,k)$ plane.
We focus on the most well-constrained eigenmode, since this is mostly $\Sigma$
with a broad oscillation in the time direction, which simplifies the discussion. In Fig.~\ref{fig:sig_mode1_lensing} we plot the $\Sigma$ part of the eigensurface for this mode, with curves of constant $l$ overlaid. Deviations from GR occur in the lensing signal because as we sum up the contributions along these curves in evaluating the integral in Eq.~\eqref{eq:limber}, we pass through regions where $\Sigma$ does not obtain its GR value. The relative importance of these regions in the above integral depends on the magnitude of the deviation in $\Sigma$, the radial distance through the region, and the product of the time-dependent geometric lensing kernel in Eq.~\eqref{eq:limber} and the scale- and  time-dependent value of $\mathcal{P}_{(\Psi+\Phi)/2}(k;\eta)$ in GR. The geometric kernel favours late times and the power spectrum favours large scales and early times.
As an example, consider $l=300$. The corresponding hyperbola in Fig.~\ref{fig:sig_mode1_lensing} passes through a region where $\Sigma < 1$, and a region at lower $\chi$ where $\Sigma>1$. The geometric lensing kernel up-weights the patch at lower $\chi$ thus raising $C_l^{dd}$ above its GR value, consistent with Fig.~\ref{fig:TT_dd_13}. As we decrease $l$, we start to pick up the large region with $\Sigma < 1$ at late times at the bottom of Fig.~\ref{fig:sig_mode1_lensing}, as well as the large region in the top left corner at large scales with $\Sigma < 1$. This reduces $C_l^{dd}$ below its GR value giving the negative feature in Fig.~\ref{fig:TT_dd_13} around $l\approx 100$. At even lower $l$, we start to accumulate relatively more of the `positive' part of the $\Sigma$ oscillation, where $\Sigma > 1$. This drives the upturn in Fig.~\ref{fig:TT_dd_13} at low $l$, although the Limber approximation become increasingly poor on these large angular scales.

In contrast, most of the well-constrained $\mu$ modes exhibit oscillations in scale, with the third most well-constrained mode having a single broad feature in the $(k,\eta)$ plane. As has been mentioned, CMB lensing brings limited scale information due to reconstruction noise, and the late-time ISW effect only affects
the large-scale CMB where cosmic variance is large. However, the 21\,cm signal can give excellent scale information, as well as time-dependence from the tomography. Figure~\ref{fig:z1z1_z3z3_13} shows the fractional difference from GR of this signal at low and high redshifts. The well-constrained $\mu$ modes thus come mainly from the 21\,cm signal, which can be confirmed by running the PCA using only 21\,cm as an observable. In that case, we
find the same $k$-space oscillations, although the overall constraints on modified gravity are poor due to degeneracies with the standard parameters which are not well constrained in the absence of CMB information.

The 21\,cm power spectrum for the best-constrained eigenmode has an \emph{excess} over GR on large angular scales at high redshift, with a magnitude larger than for the third-best constrained; see Fig.~\ref{fig:z1z1_z3z3_13}. This may seem surprising at first, since the $\mu(\eta,k)$ eigensurface for the best-constrained mode is low amplitude compared to the third mode and negative. However, the effects of $\Sigma$ are non-negligible on horizon-scale modes in the 21\,cm signal, and these purely relativistic effects drive the large deviation from GR at low $l$ for this mode. At $l=2$, roughly 75\% of this deviation can be attributed to the new terms in Eq.~\eqref{eq:perturb2} and their cross-correlation with the standard terms, the dominant contribution coming from the potential terms. The remaining 25\% comes from non-negligible relativistic terms in the dynamical equations for the density and velocity.

To understand which term in the 21\,cm signal contributes most to the $\mu$ constraints, we repeat the PCA without the density contribution. The number of principal components with eigenvalues below unity then decreases to 10. However, if we perform the PCA with bins (as in~\citep{2012PhRvD..85d3508H}), this number rises to 26, above that obtained with nodes. This reflects the fact that the redshift-space distortion term is significantly more sensitive to $\mu$, and particularly to its temporal variation $\dot{\mu}$, than the density term. Hence if there were a way to separate density from redshift-space distortion in the 21\,cm signal, this could significantly improve sensitivity to modifications of gravity, although uncertainties in the bias would propagate through to the variance of any density-free estimator.

To summarise, in models where the bias is constant and fixed, we are able to extract roughly $22$--$29$ well-constrained modes from the modified-gravity sector. Our observables are most sensitive to broad radial oscillations in $\Sigma$, and broad wavenumber oscillations in $\mu$. Addition of 21\,cm information dramatically improves the number of well-constrained modes, increasing this number from 10 to 22 using the $\sigma < 1$ threshold reference.
The results in this section assume $N_{\text{win}}=20$ windows equally spaced in frequency. We speculate that increasing the number of bins could improve constraints on the individual mode amplitudes (i.e.\ reduce their eigenvalues) by as much as a factor of $\sqrt{10}$. However, we would not expect significant changes in the form of the eigenmodes since the the 20 redshift windows already over-sample the fastest temporal variation supported by our grid. Given the sharp upturn in the spectrum of eigenvalues in Fig.~\ref{fig:eigvals_full}, the number of well-constrained modes should also not change significantly with an increase in $N_{\text{win}}$. As a concrete (and computationally tractable) example, running the PCA with 25 equally-spaced windows does not have a significant effect on the eigenvalues of the well-constrained modes, but does improve the constraints on some of the poorly-constrained modes (although the eigenvalues of these modes still remain well above unity).

Finally, the PCA provides us with a useful consistency check. We can map
$\mu$ and $\gamma$ in the $B_0$ parametrization of $f(R)$ gravity,
Eq.~\eqref{eq:fRgamma} , onto the 32 nodal values and hence constrain $B_0$
by transforming the Fisher matrix. Doing this gives a $1\sigma$ marginalised error on $B_0$ of $3.5 \times 10^{-5}$, which is within 5\% of the value found by using $B_0$ as a parameter directly.

\subsubsection{Single free bias parameter}
\label{subsec:freeb}

Treating the bias as constant and perfectly known is unrealistic. 
In this subsection, we treat the bias as constant -- a reasonable assumption on
large scales -- but unknown, and marginalise over it in the forecast about a fiducial value $b=2$.
We thus expand the dimension of our total PCA parameter space to 40.

We find that the combination of 21\,cm and CMB can constrain the bias with a marginalised $1\sigma$ error of $0.05$, i.e. to a precision of $2.5\%$. As a reference, the cross-correlation of DLAs with the Lyman-$\alpha$ forest from \citep{2012arXiv1209.4596F} measures $b = (2.17 \pm 0.20)\beta_F^{0.22}$, where $\beta_F$ is the Lyman-$\alpha$ forest redshift-space distortion parameter which is expected to be above unity. Taking $\beta_F=1$, this represents a 9\% measurement of $b$, but this ignores uncertainty in $\beta_F$ from uncertainty in the cosmology
and from modifications to gravity.

\begin{table}
  \caption{Fractional increase in $1\sigma$ marginalised errors due to marginalising over a constant bias factor.}
  \begin{center}
    \begin{tabular}{c c}
      \hline \hline
       Parameter & $\sigma$(free $b$)/$\sigma$(fixed $b$) \\
       \hline
       $\Omega_bh^2$ &  1.080\\
       $\Omega_ch^2$ &  1.584\\
       $h$ & 1.148 \\
       $\tau$ & 1.000 \\
       $A_s$ & 1.029 \\
       $n_s$ & 1.137 \\
       $\Omega_{\nu}h^2$ & 2.218 \\
       \hline \hline
    \end{tabular}
  \end{center}
  \label{table:biaschanges}
\end{table}

The inclusion of a constant bias as a free parameter has negligible effect on the modified-gravity constraints. The number of well-constrained eigenmodes in the PCA does not change. The main effect of marginalising over $b$ is a degradation in the error on $\Omega_{\nu}h^2$; see Table~\ref{table:biaschanges}, where we list the ratios of marginalised errors on cosmological parameters with and without
marginalising over bias. The scale dependent suppression of power induced by massive neutrinos discussed in Sec.~\ref{subsec:fR_results} can be (very approximately) mimicked by changing the bias. Even though we have forced $b$ to be scale-independent, increasing it is not equivalent to a scale-independent amplitude change, since the redshift-space distortion part of the brightness temperature is unaffected by bias
and is suppressed relative to the density term on small scales due to the width
of the window function.


\subsubsection{General bias function}
\label{subsec:generalb}

Although the assumption of scale-independent bias is probably safe on the scales of interest, neglecting time-dependence in the bias is again likely to be unrealistic. Indeed, simple analytic models suggest that the bias may show strong evolution with redshift \citep{2010ApJ...718..972M}. With this is in mind, we test the robustness of our constraints to more general assumptions about the bias.

Firstly, we introduce $N_{\mathrm{win}}=20$ scale-independent bias parameters, one for each of the frequency windows, and marginalise over these parameters around a
constant fiducial bias model, as in \citep{2009PhRvD..79h3513Z}. We note that the analytic models of \citep{2010ApJ...718..972M} suggest that the bias is an increasing function of redshift, so allowing the separate biases to vary independently is likely over-pessimistic. The constraint on $B_0$ degrades by a factor of 2.6, being $9.6 \times 10^{-5}$ at 68\% confidence. That we still obtain a constraint better than the CMB-alone case suggests that scale information in the 21\,cm density term, as well as information in the redshift-space distortions, are still enough to improve over Planck. However, the degradation in the constraint over the fixed-bias case suggests that the time evolution imparted in the density term in $f(R)$ models, as observed in the lowest redshift windows, carries significant weight when constraining $B_0$.

In the PCA, marginalising over 20 scale-independent bias parameters reduces the number of PCA eigenmodes with eigenvalues less than unity to 20,
i.e.\ only two modes are lost by throwing away the time-dependent information in the 21\,cm density term. The full spectrum of modes is shown in Fig.~\ref{fig:eigvals_full}. We put this encouraging result down to the significant role played by redshift-space distortions and scale-dependence in the 21\,cm density term in constraining deviations from GR. 

Finally, we consider how the constraints degrade if \emph{no} prior knowledge on the bias is assumed, thus removing all information in the 21\,cm density term.
Instead of adopting a general functional form $b(a,k)$ for the bias and marginalising, we adopt a simpler treatment, as follows, which should give results that are nearly equivalent.
When computing the Fisher matrix in Eq.~(\ref{eq:fisher_matrix}), we set the 21\,cm density term equal to zero when computing derivatives of power spectra, but retain it when computing the covariance matrix between power spectra. This amounts to throwing away information from the density power spectrum (and its cross-correlations with the other terms in the brightness temperature perturbation), but retaining its contribution to the cosmic variance. In this `worst case scenario', where no prior information on time or scale behaviour of the bias is available, we find that the constraint on $B_0$ degrades to $1.6 \times 10^{-4}$, close to the CMB-alone value
of $1.9\times 10^{-4}$ and a factor of 2.9 worse than the fixed-bias case. This suggests that most of the improvement on $B_0$ brought by 21\,cm intensity mapping is due to the density term, which is sensitive to assumptions about the bias. In the PCA, the number of well-constrained PCA eigenmodes drops to 13; see Fig.~\ref{fig:eigvals_full}. Thus, although a large proportion of the modes we unveil with intensity mapping are lost when no assumptions are made on the bias, we can still improve over the CMB alone due to the contribution of redshift-space distortions to the signal.

\section{Conclusions}
\label{sec:conc}

We have shown that a typical near-term low-budget 21\,cm intensity mapping experiment will be able to improve constraints on modified gravity theories considerably, when combined with CMB observations from Planck. In particular, this observable will improve constraints on the $B_0$ parameter of $f(R)$ gravity by a factor of 15 over current cosmological constraints \citep{PhysRevD.85.124038}.

By performing a principal component analysis on the free functions of the modified-gravity sector $\mu$ and $\Sigma$ (or alternatively $\gamma$)
in models with a $\Lambda$CDM expansion history,
we estimate that the 21\,cm experiment, combined with Planck, will provide useful information on
around 20 modes. This improves on the 10 modes forecast for Planck
alone, showing that 21\,cm intensity mapping has significant additional constraining
power in general modifications of gravity.
Although our results do not seem to be competitive with those
forecast in~\citep{2012PhRvD..85d3508H} for future galaxy clustering and
cosmic shear surveys, our analysis is more conservative since we
adopt a more realistic non-linear cut-off scale, motivated by recent
$N$-body simulations in $f(R)$ models~\cite{2012MNRAS.425.2128J}.
The 21\,cm experiment
is also much smaller scale and cheaper than galaxy imaging and spectroscopic
surveys. More generally, the astrophysical
and instrumental systematic effects are quite different and it is the combination of
all such cosmological probes of modified gravity, including their cross-correlations,
that will allow for accurate and precise constraints.

By studying the shape of the well-constrained eigenvectors of $\mu$ and $\Sigma$ in the $(\eta,k)$ plane, we can identify the `sweet-spots' of the CMB-plus-21\,cm
data combination. We find that the combination of observables is most sensitive to broad temporal variations in $\Sigma$, and broad scale variations in $\mu$. The former is probed primarily by the CMB, through the late-time ISW effect and
gravitational lensing, although 21\,cm has an important role
here in breaking degeneracies with $\mu$ that exist in a CMB-only analysis.
In the models we consider, with only late-time modifications to GR, the ISW effect
only probes large-scale modifications. Similarly, the scale information available from
CMB lensing with Planck will be limited by the high statistical noise in the lensing reconstruction, although this will improve considerably with ongoing high sensitivity
polarization measurements. In contrast, the 21\,cm survey provides
excellent scale information which allows scale-dependent modifications in
$\mu$ to be well constrained.

We have also considered the constraints that may be placed on a one-parameter model of modified gravity, specifically the $B_0$ parametrization often used to study $f(R)$ models. We find that our combination of observables should permit a 95\% upper limit
of $7\times 10^{-5}$ on $B_0$ after marginalising over cosmological parameters in flat models
with $\Lambda$CDM expansion histories.
Although this is not competitive with local constraints or forecasts from future weak lensing and galaxy surveys, it does suggest that useful constraints from cosmology can be obtained with the kind of low-budget experiments considered in this work.

We investigated the robustness of our constraints to uncertainties in the biased clustering of neutral hydrogen with respect to the underlying dark matter. When making no assumptions about the time dependence of the bias, but retaining scale-independence, we find that the 95\% upper limit on $B_0$ degrades to $2\times 10^{-4}$, but we lose only two well-constrained modes from the PCA. However, when throwing away all information from the density term, the upper limit on $B_0$ increases to $3\times 10^{-4}$, and we only constrain 13 modes. This suggests that although the scale-dependent growth imparted by modified gravity into the density term helps a lot in constraining modified gravity, the importance of redshift-space distortions in the signal is such that even without the density term we still get tighter constraints than a CMB-alone analysis.

Although we have considered a specific experimental set-up in this work, we have tried to keep the survey details as general as possible. In principle all of our survey parameters could be different, but we would not expect our conclusions to change significantly provided that the measurements are signal-dominated
at all redshifts on scales where linear perturbation theory is reliable (as they are
in our analysis) and a significant fraction of the extra-Galactic sky is surveyed.
Generally, the ratio of signal to thermal noise is best at low redshifts, and this
is also where the foregrounds are lowest.
Since any redshift bin can only experience the knock-on effects of modified growth that occur prior to that epoch, the low redshift bins carry a greater amount of information on modified gravity. It is precisely in these bins where the noise from the experiment is smallest.

One limitation of this work is our restriction to 20 narrow frequency windows equally spaced across the 400--800\,MHz range for computational reasons. This is certainly lossy, equivalent to throwing away a significant fraction of the survey volume. While our $B_0$ constraint is rather stable to increasing the number of windows or using closely spaced windows at low redshift, it is not clear how sensitive our PCA results are to the same changes. Since our most well-constrained modes exhibit broad features in redshift, we do not expect the form of these modes would change, although, generally, we expect the errors on some of the mode amplitudes would improve several-fold. The total number of well-constrained modes is expected to be stable. Ways to improve our current implementation include using all $(\nu_{\text{max}}-\nu_{\min})/\Delta \nu$ windows but only computing correlations that are expected to be significant, or, better, finding an improved radial basis that approximately decorrelates the signals. However, a detailed study of these issues is beyond the scope of this work.

A further extension to this work would be to increase the grid resolution of the PCA. Since adding more frequency windows can improve constraints on modes exhibiting grid-scale variations, we may be able to recover more well-constrained modes this way. However, it is unclear how physical rapidly varying functions in the modified-gravity sector are. We have shown that our adopted grid resolution is sufficiently fine to reproduce $f(R)$ gravity, and we speculate that the same would be true for other models. A complete Bayesian analysis would treat grids with different resolution as independent models, with the data deciding which model has the highest posterior probability, but exploring this issue is also beyond the scope of this work.

Finally, we have provided a consistent derivation of the perturbed 21\,cm brightness temperature in linear theory in any metric theory of gravity.
The new expression we derive includes all line-of-sight and relativistic effects, but these extra terms make only a small correction on large scales. The size of the correction increases with redshift and width of the frequency windows. We include the new terms in our analysis for numerical consistency, although they are negligible for our application to modified gravity. 


Our results show that 21\,cm intensity mapping provides an exciting opportunity for testing GR on cosmological scales by probing the time and scale dependence of
the clustering of neutral hydrogen. Additionally, such surveys allow geometric tests of
the expansion history through the baryon acoustic oscillation feature and weak lensing
tomography, at a fraction of the price of other large-scale structure surveys.

\section{Acknowledgments}
It is a pleasure to thank Sebastien Clesse, Pedro Ferreira and Levon Pogosian for useful discussions. We especially thank Ue-Li Pen for early collaboration on this work, and the anonymous referee for helpful suggestions which improved this paper. AH wishes to thank Philippe Brax, Elise Jennings and Richard Shaw for helpful correspondence. AH is supported by an Isaac Newton Studentship from the University of Cambridge, and the Isle of Man Government. CB is supported by the Herchel Smith Fund and by King's College Cambridge.


\bibliography{paper5_rev}

\end{document}